\documentclass[fleqn,usenatbib]{mnras}

\usepackage{newtxtext,newtxmath}
\usepackage{mathrsfs}

\usepackage[T1]{fontenc}

\DeclareRobustCommand{\VAN}[3]{#2}
\let\VANthebibliography\thebibliography
\def\thebibliography{\DeclareRobustCommand{\VAN}[3]{##3}\VANthebibliography}

\usepackage{graphicx}	
\usepackage{amsmath}	
\usepackage{ulem}
\usepackage{multirow}
\usepackage{threeparttable}

\newcommand{\beq}               {\begin{equation}}
\newcommand{\eeq}               {\end{equation}}

\newcommand{\peryr}             {\,{\rm yr}^{-1}}
\newcommand{\Gyr}               {\,{\rm Gyr}}
\newcommand{\Kalvin}            {\,{\rm K}}

\newcommand{\Mpc}               {\,{\rm Mpc}}
\newcommand{\Msun}              {\,{\rm M}_\odot}
\newcommand{\kmsMpc}            {\,\,{\rm km}\,\,{\rm s}^{-1}\,{\rm Mpc^{-1}}}

\newcommand{\sls}               {\slshape}
\newcommand{\YunnanII}          {{\sc yunnan-ii}}
\newcommand{\Yehb}              {{\slshape YII-ehb}}
\newcommand{\Ysin}              {{\slshape YII-sin}}
\newcommand{\Ybin}              {{\slshape YII-bin}}

\newcommand{\GABE}              {{\sc gabe}}

\newcommand{\Om}                {\Omega_{\rm m}}
\newcommand{\Ob}                {\Omega_{\rm b}}
\newcommand{\OL}                {\Omega_{\rm \Lambda}}
\newcommand{\seight}            {\sigma_{\rm 8}}
\newcommand{\Hzero}             {H_{\rm 0}}

\newcommand{\Mstar}             {M_{\rm *}}
\newcommand{\Zstar}             {Z_{\rm *}}
\newcommand{\Zsun}              {Z_\odot}
\newcommand{\Fy}                {F_{\rm y}}
\newcommand{\Fb}                {F_{\rm b}}
\newcommand{\Eb}                {E_{\rm b}}
\newcommand{\mpri}              {m_{\rm 1}}
\newcommand{\msec}              {m_{\rm 2}}

\newcommand{{\UVXintrinsic}}    {UV-upturn\textsubscript{intrinsic}}
\newcommand{{\UVXdust}}         {UV-upturn\textsubscript{dust}}

\title[UV-upturn galaxies in semi-analytic models]{Two categories of UV-upturn galaxies revealed by semi-analytic models}

\author[Z. Jiang et al.]{
\parbox{\textwidth}{Zhen Jiang,$^{1,2}$\thanks{E-mail: jiangzhen@nao.cas.cn}
Cheng Li,$^{1}$\thanks{E-mail: cli2015@tsinghua.edu.cn}
Fenghui Zhang,$^{3,4,5}$
and Shuang Zhou$^{6}$}\vspace{0.4cm}
\\
$^{1}$Department of Astronomy, Tsinghua University, Beijing 100084, China\\
$^{2}$National Astronomical Observatories, Chinese Academy of Sciences, Beijing 100101, China\\
$^{3}$Yunnan Observatories, Chinese Academy of Sciences, Kunming, 650216, China\\
$^{4}$Key Laboratory for the Structure and Evolution of Celestial Objects, Chinese Academy of Sciences, Kunming, 650216, China\\
$^{5}$International Centre of Supernovae, Yunnan Key Laboratory, Kunming 650216, P. R. China\\
$^{6}$INAF–Osservatorio Astronomico di Brera, via Brera, 28, 20159 Milano, Italy\\
}

\pubyear{2025}

\begin{document}
\label{firstpage}
\pagerange{\pageref{firstpage}--\pageref{lastpage}}
\maketitle

\begin{abstract}
UV-upturn galaxies are characterized by unusually excessive flux in the far-ultraviolet (FUV) band, observed in some elliptical galaxies and the bulges of disk galaxies. We examine UV-upturn galaxies within the semi-analytic model {\GABE}, which embeds the formation of extreme horizontal branch (EHB) stars—proposed as key candidates responsible for the UV-upturn phenomenon. We have analyzed all related physical processes, including stellar evolution, initial mass functions (IMFs), dust attenuation, galaxy age, metallicity, and binary fractions, in an effort to determine which processes play significant roles. Our findings reveal two categories of UV-upturn galaxies in the semi-analytic model, each with distinct formation channels: old metal-rich quenched elliptical galaxies, which are intrinsic UV-upturn galaxies induced by EHB stars within their old stellar populations, and dusty star-forming galaxies, which are relatively young and may also be photometrically identified as UV-upturn galaxies when accounting for dust attenuation. Dust attenuation contributes to $20\%-60\%$ of the UV-upturn galaxies, depending on the specific dust attenuation models adopted. With the binary star formation model of EHB stars, both of these formation channels exhibit strong preferences for high stellar metallicity. The high-mass end slope of the IMFs is found to have a marginal effect, indicating that a universal IMF is adequate for studying the UV-upturn phenomenon.
\end{abstract}



\section{Introduction}
\label{section:intro}

The ``UV upturn'' or ``UV excess'' is an unusual phenomenon characterized by excessive far-ultraviolet (FUV) flux observed in some elliptical galaxies and the central bulge of some spiral galaxies, such as M31 \citep[e.g.][]{Rosenfield12}. This phenomenon was discovered with the first FUV observations by the {\sls Orbiting Astronomical Observatory 2} \citep[{\sls OAO-2};][]{CODE69,CODE79}. Early imaging and spectroscopic studies found that the UV excess is spatially extended \citep[e.g.][]{Deharveng76,Deharveng80,Deharveng82}, following the profile of visible light \citep[e.g.][]{Welch82,Bohlin85}, but lacking association with massive stars \citep[e.g.][]{King92,Bertola95} or star formation activities \citep[e.g.][]{OConnell92}. Further spectral and photometric analyses have shown that white dwarfs alone cannot account for the entire UV-upturn phenomenon \citep{Werle20}. These observations collectively suggest that the UV excess is not due to young massive stars, but rather originates from old stellar populations other than white dwarfs. Additionally, studies examining UV-upturn galaxies across various low-redshift environments have found no dependence of UV excess on cluster environment, group multiplicity, or satellite velocity dispersion. This suggests that the UV upturn is driven by processes internal to the galaxies themselves \citep[e.g.][]{Yi11, Loubser11, Boissier18, Ali18c, Ali19environment, Phillipps20}.

Studies of galaxies at higher redshifts generally report a moderate to negligible evolution in UV excess over redshifts from $z \sim 0$ up to $z \sim 0.6-0.7$ \citep[e.g.][]{Brown98, Brown00redshift, Brown03, Ree07, Donahue10, Boissier18, Atlee09}. Beyond this range, the occurrence of the UV-upturn phenomenon declines rapidly \citep[e.g.][]{LeCras16, Ali18b, Ali18c, Ali21, DePropris22, Ali24}. Although \cite{LeCras16} observed the near disappearance of the UV-upturn phenomenon at $z \sim 1$ in a sample of massive galaxies from the Baryon Oscillation Spectroscopic Survey \citep[BOSS,][]{BOSS}, \cite{Lonoce20} reported detecting UV-upturn galaxies at $z \sim 1.4$ in the Cosmological Evolution Survey (COSMOS) field. In this context, the UV-line indices could only be explained after considering an old, UV-bright stellar population, consistent with the previously identified origin from old populations in low-redshift UV-upturn galaxies.

Another well-known observational result is the metallicity dependence of the UV upturn, originally identified by \cite{Burstein88} through the study of the {\sls International Ultraviolet Explorer} ({\sls IUE}) spectra of 24 quiescent early-type galaxies (ETGs). They found that the FUV-V color of these galaxies became bluer with a higher Lick ${\rm Mg_2}$ absorption index, indicating higher metallicity. While this correlation has been re-evaluated with subsequent observational data, early studies reported either no correlation \citep[e.g.][]{Ohl98, Deharveng02, Rich05} or only modest correlations \citep[e.g.][]{Boselli05, Donas07}. More recent research has accumulated evidence supporting this correlation. For example, \cite{Bureau11} demonstrated strong correlations between the FUV-V and FUV-NUV colors and the integrated Mg$b$ index in nearby ETGs with purely old stellar populations using FUV/NUV photometry from {\sls GALEX} \citep{GALEX} and integral field spectroscopy from SAURON \citep{SAURON_integral_filed}. Additionally, \cite{Jeong12} found that the global correlation of the FUV excess with metallicity also applies at a local level. Furthermore, \cite{Carter11} observed that the strength of the UV upturn correlates more strongly with [$\alpha$/Fe] than with [Z/Fe]. The correlation between FUV-NUV and FUV-optical colors with metallicity is also observed in rich galaxy clusters in the local Universe, such as Coma \citep{Smith12} and Virgo \citep{Akhil24}.

Theoretically, extreme horizontal branch (EHB) stars and their hot post-HB descendants, including post-early asymptotic giant branch (AGB) stars and AGB-manqu\'e stars, have been proposed as key candidates responsible for the UV excess in old, long-lived stellar populations \citep{GR90, GR99}. EHB stars are core-helium-burning stars characterized by extremely thin hydrogen envelopes ($M_{\rm env} < 0.02\Msun$), with temperatures $T_{\rm e} \sim 25,000\Kalvin$ and masses of approximately $0.5\Msun$. They fall below the main sequence of massive stars on the Hertzsprung-Russell (H-R) diagram and are particularly energetic in the FUV band. Given that $10-20\%$ of evolving stars pass through the EHB channels, the observed FUV luminosities can be reasonably explained \citep{GR99, Brown95}. The detection of EHB and post-EHB stars in M32 by the {\sls Hubble Space Telescope} provided direct evidence for the EHB origin of the UV excess \citep{Brown00HB, Brown08}; for a comprehensive review, see \cite{OConnell99}. We note that, in this work, we do not meticulously distinguish the effect of EHB stars with that of their descendants. Hereafter the term ``EHB stars'' represents the sum of them for simplicity.

Although the EHB origin has become a consensus among the community, the specific physical mechanisms driving the formation of the thin hydrogen envelope—critical for EHB star formation—remain unclear \citep{Iben70, Dorman92}. Several theoretical models have been proposed to explain the formation of this thin envelope. For instance, the ``metal-poor single star model'' interprets EHB stars as products of the metal-poor tail of a stellar population with a wide metallicity distribution \citep{Lee94, Park97}. However, this model requires the stellar population to be extremely old ($18-20\Gyr$), even older than the age of the Universe. In contrast, the ``metal-rich single star model'' assumes that EHB stars originate from metal-rich stars that undergo significant envelope loss near the tip of the red giant branch \citep[RGB, e.g.][]{Bressan94, Dorman95, DCruz96, Yi97, M05}. This model posits a relatively large mass-loss rate that increases with metallicity during the RGB phase in order to reduce the envelope mass \citep{Yi98}. A more recent model, the ``He-rich single star model'' \citep[e.g.][]{Chung11, Chung17}, suggests that EHB stars form from subpopulations with high initial helium abundances, which possess high mean molecular masses and short lifetimes, leading to low turn-over masses and thin envelopes. Additionally, a higher abundance of helium results in HB stars burning more of their hydrogen envelope during their HB phase \citep{Horch92, Dorman95}. Both the metal-rich and He-rich single star models predict a relatively strong evolution of the UV excess phenomenon with both metallicity and age of the stellar population, which can account for the observed correlation noted by \cite{Burstein88} as well as the redshift evolution described by \cite{Ali21, Ali24}. The He-rich model effectively reproduces the He-rich EHB subpopulations observed in globular clusters (GCs) \citep[][]{Chung11, Peacock18}, while the increased mass-loss rate during the RGB phase required by the metal-rich model has not been observed in star clusters \citep{Miglio12, McDonald15, Williams18}. Given these observations, the He-rich model is favored over the metal-rich model, assuming that the EHB stars in star clusters share the same origin as those in ETGs \citep{Goudfrooij18}.

In contrast to the single star models, the ``binary star model'' considers various binary interactions as the formation mechanism for EHB stars. An EHB star may form after the common-envelope phase of a close binary system if helium is subsequently ignited in the star's core. Conversely, if mass transfer remains dynamically stable, the star's envelope will be stripped via stable Roche lobe overflow (RLOF), resulting in a remnant core with a long orbital period that may also evolve into an EHB star if helium in the core is ignited. Furthermore, a single EHB star can originate as the merger remnant of two helium white dwarfs, provided the total mass is less than the Chandrasekhar limit \citep{Webbink84, Iben86, Han98}. \cite{Han02, Han03} meticulously examined all these formation channels and conducted a binary population synthesis study of EHB stars, successfully explaining the observed EHB star distributions without making specific assumptions regarding mass-loss rates and He abundance. \cite{Han07} further developed an evolutionary population synthesis (EPS) model. In this model, for old stellar populations, the FUV-NUV color decreases substantially with age, while the FUV-$r$ color exhibits a relatively weak correlation with age, consistent with the observed weak evolution of the FUV-$r$ color at $z < 0.6$. Although only solar metallicity was considered, the authors argued that the UV excess phenomenon should not depend strongly on metallicity, as the mass transfer processes related to the formation of EHB stars are primarily influenced by initial orbital parameters rather than metallicity. Consequently, the binary star model predicts a weak or negligible correlation between the FUV-$r$ color and metallicity, which contradicts previous observational studies.\footnote{We note that the metallicity dependence of the UV upturn in binary star models is revisited in this work, leading to somewhat different conclusions. Details can be found in~\autoref{section:discussion-binary}.}

In previous studies, the evolution and correlation of the UV-upturn phenomenon were generally derived for simple stellar populations (SSPs), where stars were assumed to form through a singular starburst \cite[e.g.][]{HB13, HB14}, composite stellar populations (CSPs) comprising two SSPs \citep[e.g.][]{Han07}, or CSPs that incorporated chemical evolution histories \citep[e.g.][]{Ali21}. Assumptions and simplifications were applied to the star formation histories (SFHs) of ETGs. In this work, we examine the formation and evolution of the UV-upturn phenomenon at galactic scales (as opposed to stellar population scales), particularly including variations in SFHs by embedding the formation model of EHB stars within a cosmological framework provided by semi-analytic models (SAMs). In SAMs, the SFHs of galaxies are generated self-consistently through a series of galactic physical processes described by simplified models or empirical relations. Combined with N-body merger trees of dark matter haloes and tracking the formation and evolution of galaxies within these haloes, SAMs have successfully reproduced a substantial amount of observational data at both low and high redshifts \citep[e.g.][]{White91, Kauffmann99, Croton06, DeLucia07, Guo11, Henriques15, Henriques20}. The SAM employed in this study is {\GABE} \citep[Galaxy Assembly with Binary Evolution,][]{Jiang19}, which includes a comprehensive set of galaxy formation recipes, encompassing reionization, gas cooling, star formation, supernova feedback, black hole growth, AGN feedback, galaxy mergers, and more. Notably, the evolution of binary stars, particularly the channels for the formation of EHB stars, is integrated using the {\YunnanII} EPS model \citep{Zhang10}.

In addition, the approach of SAMs enables us to examine the influence of various galactic physical processes on the UV-upturn phenomenon. In particular, we will consider stellar initial mass functions (IMFs) with varying slopes at the high-mass end, dust attenuation curves with varying slopes and UV bump strengths, and the fractions of binary stellar populations and young stellar populations. By incorporating these factors into SAMs and comparing the predicted occurrence rates of UV-upturn galaxies, we are able to determine which of these factors might exert important influences on the identification and formation of UV-upturn galaxies. 

The organization of this paper is as follows. In~\autoref{section:method}, we briefly introduce the adopted semi-analytic model. In~\autoref{section:results}, we demonstrate properties of modeled galaxies, and conduct a thorough examination of related physical processes to assess their influence on the formation of UV-upturn galaxies, including various EPS models, IMFs, dust attenuation, and the joint dependencies. We discuss the dependence of UV upturn in binary star model and the role of dust attenuation in~\autoref{section:discussion}, and summarize our conclusions in~\autoref{section:conclusion}.

\section{Method}
\label{section:method}

\subsection{Semi-analytic Model}

We utilize the semi-analytic model (SAM) {\GABE} (Galaxy Assembly with Binary Evolution) developed by \citet{Jiang19} as the basic tool for our study. The {\GABE} model is implemented using the Millennium Simulation \citep{Springel05}, a dark matter-only cosmological simulation which adopts the cosmological parameters ($\Om = 0.25$, $\Ob = 0.045$, $\OL = 0.75$, $n = 1$, $\seight = 0.9$ and $\Hzero = 73 \kmsMpc$) derived from a combined analysis of the 2dFGRS \citep{Colless01} and the first-year WMAP data \citep{Spergel03}. Dark matter haloes and subhaloes are identified with a friends-of-friends group finder \citep{Davis85} and {\sc subfind} \citep{Springel01}, respectively. The merger trees are constructed by tracing the formation and merger history of each halo and subhalo with the D-Tree algorithm \citep{Jiang14}, upon which {\GABE} is applied. The simulation box size is $685\Mpc$, which is large enough to suppress cosmic variance, thereby enabling reliable statistics \citep{Wang17}. The mass resolution of dark matter particles is $1.2 \times10^9 \Msun$, allowing {\GABE} to generate a complete galaxy catalogue for galaxies more massive than $\sim10^{9}\Msun$. Like other SAMs, {\GABE} includes a full set of galaxy formation recipes, including reionization, gas cooling, star formation, supernova feedback, black hole growth, AGN feedback, galaxy mergers, etc. As shown in \citet{Jiang19}, {\GABE} successfully reproduced the stellar mass function of galaxies as well as the scaling relations of galaxy properties such as metallicity, black hole mass, galaxy size and so on. The reader is referred to \cite{Jiang19} for further details of the model.

As mentioned, UV excess is observed in ellipticals or galactic bulges. In order to facilitate a more reasonable comparison between model and observation, we only consider the ``bulge region'' of each model galaxy. The ``bulge region'' of a galaxy is defined as the area within twice the half-mass radius of the stellar bulge, which is formed mainly through galaxy mergers in {\GABE}. Assuming the density profile of the galactic disk to be exponential, we calculate the volume fraction of this bulge region over the whole disk and assign the corresponding fraction of disk component to the bulge. In this way, we will compare different properties of the bulge region of model galaxies with observations, and the model galaxies mentioned hereafter are represented by their bulge regions. Note that a model galaxy defined in this way is naturally a composite of a bulge with or without a disk. The ones without disks are analogous to fully quenched elliptical galaxies in observations, while those with disks can be compared with elliptical galaxies with certain levels of star formation activities or with the bulges of spiral galaxies in observations. We only consider modeled galaxies whose bulge region is more massive than $10^{9.5}\Msun$, to reflect the fact that UV-upturn galaxies are found mainly in massive galaxies \citep[e.g.][]{LeCras16,Akhil24,Martocchia25} as well as to match the stellar mass lower limit we apply on the observational sample in~\autoref{section:colors}.

When a star formation event happens in a galaxy, triggered by disk instability or galaxy merger, a corresponding SSP is generated to record the time and strength of this event. The star formation history (SFH) of the galaxy is constructed by aggregating all SSPs throughout its history. UV and optical magnitudes are then derived for the galaxy by convolving the SFH with the photometric evolution of SSPs provided by the adopted EPS models. For the calculation of magnitudes, it is crucial to include dust attenuation,  which will be discussed in detail in~\autoref{section:dust}. Compared with previous SAMs, {\GABE} for the first time incorporated the modeling of binary star evolution by adopting the {\YunnanII}\footnote{\url{http://users.ynao.ac.cn/~zhangfh/}} EPS model \citep{Zhang04,Zhang05,Zhang10,Zhang20}. This model is constructed based on the rapid binary star evolution (BSE) algorithm of \cite{Hurley02}, which encompasses various binary interactions, such as mass transfer, mass accretion, common-envelope evolution, collisions, supernova kicks, tidal evolution, and angular momentum loss via gravitational waves. More technical details on the {\YunnanII} model, including initial distributions and spectral library, are available in \cite{Zhang20}. In \cite{Zhang10}, different formation channels of EHB stars of \cite{Han02,Han03} were also included by directly integrating the EPS model with those of \cite{Han07}, under the assumption that metallicity does not play an essential role in the formation of EHB stars. Consequently, there are three versions of the {\YunnanII} model: the single star version by deactivating all binary interactions ({\Ysin}); the binary version without the formation of EHB stars through double He white dwarfs merger, stable RLOF and common-envelope ejection under the assumption of sub-He core ignition\footnote{The minimum core mass for helium ignition could be lowered if the mass-loss rate during RGB phase is high enough, which increases the formation efficiency of EHB stars \citep{DCruz96,Han02}. We refer to such mechanism as ``sub-He core ignition''.} ({\Ybin}); and the binary version with all the above binary formation channels of EHB stars ({\Yehb}). The stellar initial mass function (IMF) of \cite{MS79} is adopted as the fiducial IMF in {\YunnanII} models, but a variety of IMF models will be considered in~\autoref{section:IMF}.

In addition to {\YunnanII} models, we also consider the EPS model of Binary Population and Spectral Synthesis \citep[{\sc bpass}\footnote{\url{https://bpass.auckland.ac.nz/}};][]{Eldridge08,Eldridge09,Eldridge17}, as well as {\sc galaxev}, the commonly-used EPS model of \citet{BC03}. The {\sc galaxev} adopts the IMF of \citet{Chabrier03} and has no formation channels of EHB stars and their descendants. The {\sc bpass} model is based on a custom version of the Cambridge stellar evolution code {\sc stars} \citep{Eggleton71} and the binary evolutionary scheme of \cite{Hurley02}. This model was initially developed to investigate the effects of massive binaries on young stellar populations and has been widely used by both stellar and extragalactic studies \citep[e.g.][]{Blagorodnova17,Ma16}. The properties of old stellar populations were also reevaluated in \cite{Stanway18}, leading to an improved concordance with observed colors and spectral indices in both globular clusters and quiescent galaxies. However, a discussion on the UV-upturn phenomena in these models was absent, and the formation channels of EHB stars through binary interactions of \cite{Han02,Han03} were not explicitly incorporated. The {\sc bpass} v2.2 model provides predictions for both single and binary populations with various IMFs.

\subsection{Definition of ``UV-upturn''}
\label{section:definition}

\begin{figure*}
  \centering
  \includegraphics[width=0.95\textwidth] {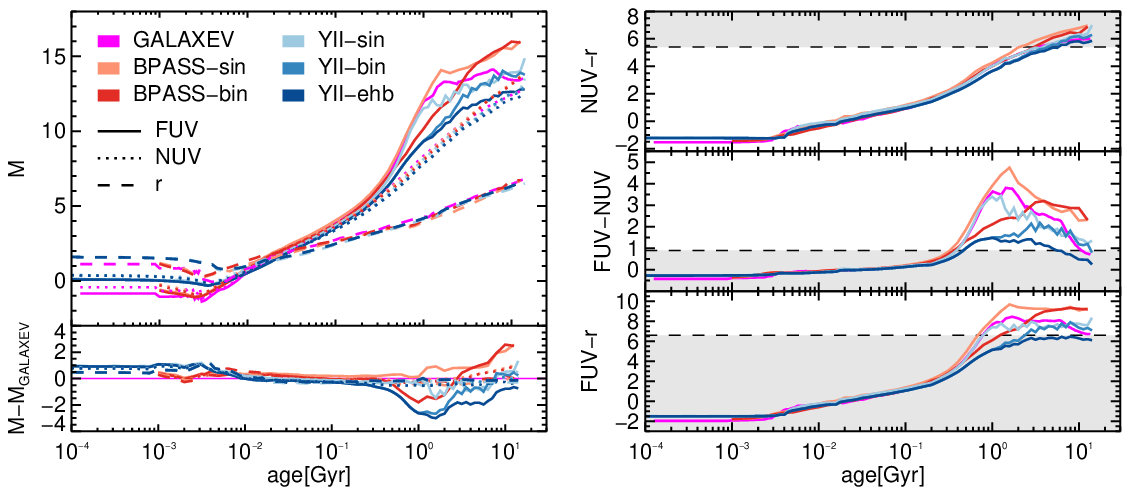}
  \caption{{\it Left:} Evolution of the FUV, NUV and $r$ band absolute magnitude of a $1 \Msun$ SSP with solar metallicity in different EPS models. For models other than {\sc galaxev}, the relative differences with respect to {\sc galaxev} are shown in the lower panel. The thin magenta horizontal line indicates zero. {\it Right:} Evolution of NUV-$r$, FUV-NUV and FUV-$r$ colors for the same models as in the left panel. The shaded regions indicate the colors expected for UV-upturn galaxies.}
  \label{fig:SSP_solar}
\end{figure*}

In this work, we adopt the UV-upturn classification scheme of \cite{Yi11}, in which a UV-upturn galaxy is defined to have NUV-$r > 5.4$, FUV-NUV $< 0.9$, and FUV-$r<6.6$. These photometric criteria were designed to select quenched galaxies with a reliable upturn feature:
\begin{itemize}
    \item NUV-$r>5.4$ is adopted to seperate  fully-quenched galaxies from those exhibiting relatively prominent residual star formation (RSF) activities. As shown by previous studies of {\sls GALEX} observations, the NUV-$r$ color presents a bimodal distribution and is a good indicator of RSF activities in galaxies \citep[e.g.][]{Kaviraj07,Salim07,Salim14}. This criterion was originally proposed by \cite{Yi05}, and the exact value adopted varies from $5.0$ to $5.5$ (e.g. 5.0, \citealt{Yi05}, \citealt{Jeong09}, \citealt{Salim14}, \citealt{Crossett17}; 5.2, \citealt{Phillipps20}; 5.4, \citealt{Schawinski07}, \citealt{Yi11}; 5.5, \citealt{Kaviraj07}). Here we use a relatively strict value of $5.4$, to better focus on quenched galaxies. It is noteworthy that even though galaxies with prominent RSF activities can be removed by this single color cut, the ones with arbitrarily small amounts of RSF can never be totally ruled out, as discussed in \cite{Phillipps20}.
    \item FUV-NUV $< 0.9$ is a criterion to recognize the upturn feature towards shorter wavelengths, based on the fact that a flat UV spectrum $F(\lambda)$ typically has a FUV-NUV color of 0.9 \citep{Yi11}. However, as suggested in \cite{Smith12} and \cite{Phillipps20}, the UV properties of quenched galaxies actually behave in a ``continuous'' way and their FUV-NUV colors span a relatively wide range \citep[e.g. 0.7-2.1 in][]{Phillipps20}. Besides, UV SEDs of different quenched ETGs or different regions within ETGs are found to be better fitted by hot components with different temperatures (e.g. $10,000-21,000\Kalvin$ in \citealt{Ali18a}; $25,000-40,000\Kalvin$ in \citealt{Martocchia25}), rather than a single temperature; and even some of the FUV-NUV$>0.9$ galaxies prefer a model with an extra hot component, e.g. M86 in \cite{Martocchia25}. These results suggest that the mechanism which induces the upturn phenomenon can not be easily picked out by a single color cut. Such a ``continuous'' behavior also presents in this work (check the ranges of FUV-NUV colors of old stellar populations in~\autoref{fig:SSP_solar} and~\autoref{fig:SSP_metal}). Nevertheless, we adopt the FUV-NUV $< 0.9$ criterion in this work to focus on the most extreme part of this continuum -- quenched galaxies which indeed have an upturn feature. Note that the value of 0.9 for FUV-NUV may slightly vary depending on the EPS model assumed. As this threshold is used to select a subpopulation within a continuous distribution, its tiny variance will have no influence on our conclusions.
    \item FUV-$r$ $< 6.6$ reflects the relatively high flux ratio between FUV to optical bands, as observed in UV-upturn galaxies. For galaxies without RSF activities, the FUV flux is a better representative of the hot evolved component than the NUV flux, which is additionally contributed by metal line blanketing from the main sequence turn-off (MSTO) stars \citep[e.g.][]{Donas07,Smith12,Martocchia25}. Therefore, FUV-$r$ color works here as a pure comparison between hot evolved stars and normal cool old stars. Furthermore, by demanding a relatively strong FUV emission, the reliability of the selected upturn feature can be ensured. The value of 6.6 was calculated for a typical quenched galaxy with FUV-NUV$=0.9$ in \cite{Yi11}, and it is also model dependent. Our test analysis shows that this model dependence has no significant influence on our conclusions, though.
\end{itemize}

Therefore, the UV-upturn classification scheme of \cite{Yi11} as described above is an empirical, model-dependent, and relatively strict definition. We adopt this definition to focus on the most typical UV-upturn galaxies, where the mechanism behind the UV-upturn phenomenon plays a more influential role. In fact, the definition or measurement of the UV-upturn phenomenon varies from one paper to another. Different definitions may result in different samples and lead to different results. In~\autoref{sec:discussion_definiton}, we will come back to this point and discuss on the influence of the definition of UV-upturn on our results.

\section{Results}
\label{section:results}

\subsection{Colors of simple stellar populations}
\label{section:EPS}

We first examine the magnitudes and color indices in UV and optical bands for Simple Stellar Populations (SSPs) across different Evolutionary Population Synthesis (EPS) models. The left panels in~\autoref{fig:SSP_solar} illustrate the evolution of absolute magnitude in the Far UV (FUV), Near UV (NUV), and $r$ band for a $1 \Msun$ SSP with solar metallicity ($Z=0.02$), for all six EPS models. The models show negligible differences in both NUV and $r$ magnitude when stellar age exceeds $\sim0.01$ Gyr. In contrast, significant variations among the models are observed in the FUV magnitude for SSPs older than $\sim0.3$ Gyr, where {\Ybin} and {\Yehb} models are more luminous than {\Ysin} and {\sc galaxev} in the FUV band, attributable to the increasing impact of binary interactions. These interactions, such as mass transfer and merger, give rise to hotter stellar objects, predominantly hot subdwarfs and blue stragglers\footnote{Blue stragglers are stars lying on the main sequence in the color-magnitude diagram of globular clusters while beyond the turning-point \citep{Sandage53}. They are believed to be produced from collisions between low-mass stars and mass transfer in close binaries \citep[e.g.,][]{Pols94,Chen04,Hurley05}.}, enhancing the FUV luminosity around $1\Gyr$ compared to single star scenarios. At ages larger than $\sim 3$ Gyr, Extreme Horizontal Branch (EHB) stars produced through the three formation channels of \cite{Han02,Han03} play a more significant role, lowering the FUV magnitude by an additional magnitude, potentially accounting for the UV-upturn phenomenon observed in elliptical galaxies, as discussed in \cite{Han07}. The {\sc bpass} models share a similar evolution with that of {\YunnanII}, but are about $1-2$ magnitudes fainter in the FUV band at large ages. Due to the lack of formation channels of EHB stars of \cite{Han02,Han03}, the single and binary models of {\sc bpass} gradually converge with each other at $\sim3$ Gyr.

\begin{figure*}
  \centering
  \includegraphics[width=1.0\textwidth] {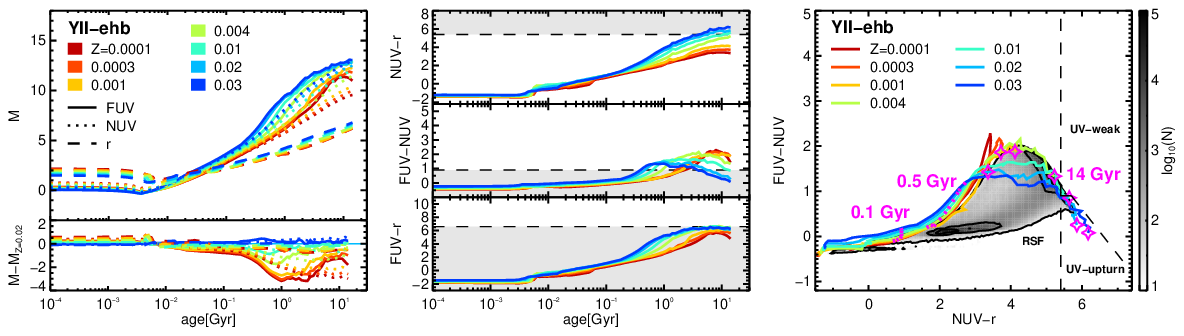}
  \caption{The left two panels are the same as in Fig.~\ref{fig:SSP_solar}, but for SSPs with different metallicities in {\Yehb} model. The right panel shows the evolutionary trajectories of SSPs with different metallicities on the color-color diagram in {\Yehb} model. Magenta dotted lines are the isochrones and magenta stars mark the evolutionary end points for each trajectories. Gray-filled region shows the number of modeled galaxies without dust attenuation in each cell at $z\sim0$. Black contours enclose $50\%$, $80\%$ and $99.8\%$ of all the modeled galaxies respectively outwards. Dashed lines are the divisions of three kinds of galaxies as proposed in \protect\cite{Yi11}: UV-upturn galaxies, UV-weak galaxies, and residual star formation (RSF) galaxies.}
  \label{fig:SSP_metal}
\end{figure*}

The right panels of~\autoref{fig:SSP_solar} illustrate the evolution of NUV-$r$, FUV-NUV, and FUV-$r$ colors for the same models, along with the UV-upturn classification scheme of \cite{Yi11}, which is NUV-$r > 5.4$, FUV-NUV $< 0.9$, and FUV-$r<6.6$, indicated by the dashed lines and shaded regions. Compared to single star models, those incorporating binary interactions consistently produce bluer SSPs. Among the color indices, NUV-$r$ exhibits minimal influence from binary interactions and shows the least variation across different EPS models. This can be attributed to the weak model-dependence of both NUV and $r$ magnitudes observed in the left panels.
The other two color indices, both involving FUV, present significant variations among the EPS models for ages exceeding $\sim0.3$ Gyr. In these cases, SSPs from single star models rapidly become red, while those from binary models remain blue due to increasing contributions from blue stragglers and EHB stars at 1 Gyr. The {\Yehb} model presents the bluest colors at large ages, with FUV-NUV progressively becoming bluer after $\sim1$ Gyr and FUV-$r$ remaining largely constant.
Notably, only the {\Yehb} model falls into the UV-upturn region when the SSPs are older than $\sim6$ Gyr. Other EPS models exhibit relatively weaker FUV flux, thereby failing to meet the criteria for FUV-NUV and FUV-$r$ colors, despite their FUV-NUV colors showing a similar trend towards the blue end at old ages. Furthermore, it is evident that young SSPs are unable to meet the UV-upturn criterion, as their relative flux of NUV over optical is excessively high.

In~\autoref{fig:SSP_metal}, we examine the dependence of magnitudes and colors on stellar metallicity, focusing solely on the {\Yehb} model for simplicity and clarity. Examination of other EPS models reveals similar trends. As observed, in all bands, SSPs of higher metallicities are brighter initially but become fainter when the stellar age exceeds $\sim0.01$ Gyr. The metallicity dependence at old ages is more pronounced in FUV and NUV than in the $r$ band, with the disparity in UV magnitudes between the most metal-rich and most metal-poor SSPs reaching up to 4 magnitudes, compared to a difference of less than one magnitude in the $r$ band.
At late ages, metal-rich SSPs are redder in both NUV-$r$ and FUV-$r$ than metal-poor SSPs. The situation is more complex for FUV-NUV, which increases to a peak before declining again, with the peak color occurring earlier for SSPs of higher metallicities (e.g., $0.6-0.7$ Gyr for $Z=0.03$ and $6-7$ Gyr for $Z=0.0001$). Consequently, only for old metal-rich SSPs with $Z>0.01$ and ages $\gtrsim4$ Gyr can both the NUV-$r$ and FUV-NUV colors evolve into the UV-upturn region, as indicated by the shaded areas in the middle panel of~\autoref{fig:SSP_metal}.

Based on these results, we expect the UV-upturn phenomenon to be associated with SSPs of high metallicities and old ages in the {\Yehb} model. This is more clearly illustrated in the right panel of~\autoref{fig:SSP_metal}, where the evolution of the {\Yehb} SSPs is plotted on the FUV-NUV versus NUV-$r$ diagram. The different solid lines indicate the evolutionary trajectories of SSPs with varying metallicities, while magenta dotted lines represent isochrones at $0.1$ and $0.5\Gyr$. For comparison, the distribution of all model galaxies {\sls without} dust attenuation is plotted in the background as gray-filled areas. The corresponding black contours outline the star-forming/blue population and quenched/red population. We find that the quenched/red population of galaxies aligns closely with "old" SSPs ($>10$ Gyr). Magenta stars denote the endpoints ($14$ Gyr) of each trajectory, showing a clear dependence on metallicity. Old and metal-rich SSPs exhibit relatively redder NUV-$r$ colors and bluer FUV-NUV colors compared to their metal-poor counterparts, with the most metal-rich SSPs ending right within the UV-upturn region. This further suggests a correlation between the UV-upturn phenomenon and both metallicity and age within the {\Yehb} model.

\begin{figure*}
\centering
\includegraphics[width=1.0\textwidth]{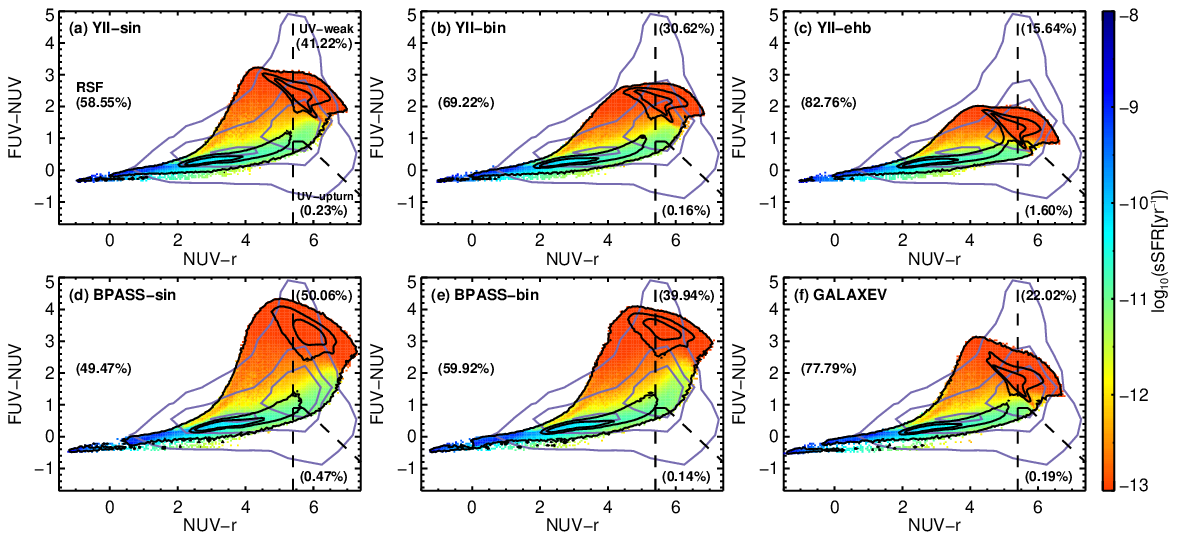}
\caption{Distributions of model galaxies on the color-color diagram, run with different EPS models. Black contours enclose $50\%$, $80\%$ and $99.8\%$ of all the modeled galaxies respectively outwards. Filled colors represent the mean specific star formation rate (sSFR) in each cell for cells with more than five galaxies. Dashed lines are the divisions of three kinds of galaxies, with number fractions listed in brackets. Purple contours show the number distributions of galaxies for the observational sample of NSA, enclosing $50\%$, $80\%$ and $97\%$ respectively outwards.}
\label{fig:SAM}
\end{figure*}

\begin{figure*}
  \centering
   \includegraphics[width=0.85\textwidth]{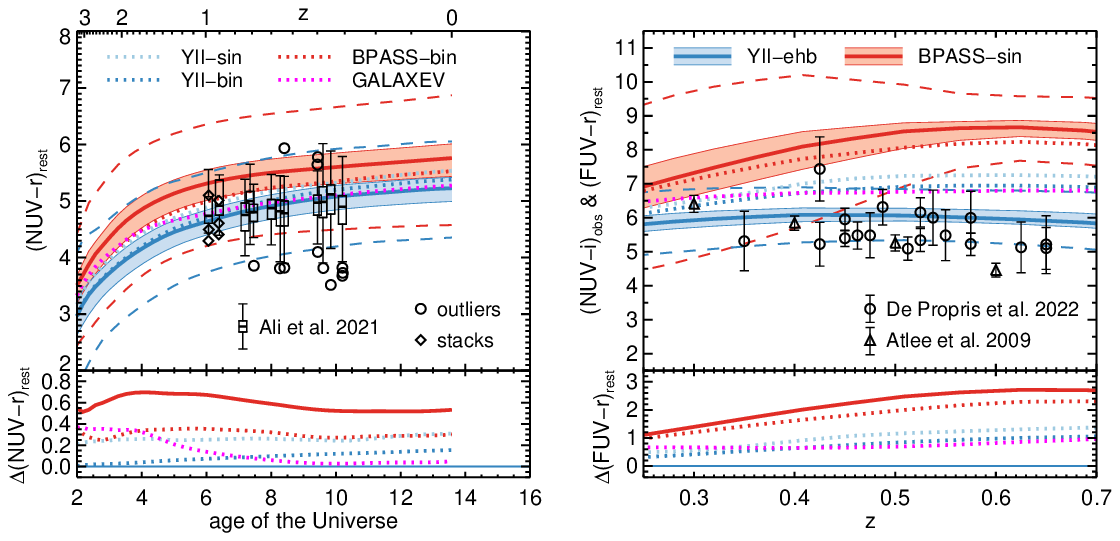}
   \caption{{\it Left:} Redshift evolution of rest-frame NUV-$r$ colors of modeled quenched galaxies (sSFR $<10^{-12}\peryr$) run with various EPS models as shown in the legend. Thick solid lines and dotted lines are the median values, and shaded regions indicate the $25\%$ and $75\%$ quartiles. Dashed lines mark the range corresponding to an additional 1.5 times the interquartile range. The {\Yehb} and {\sls BPASS-sin} models are fully illustrated as thick solid lines with corresponding distributions as the lower and upper limits of model predictions, respectively; and the median values of other EPS models are marked as dotted lines for clarity. Box plots are the observational results of ETGs in 12 clusters in \protect\cite{Ali21}, whose boxes and whiskers share the same definition as our shaded regions and dashed lines, respectively. Circles are observed outliers beyond the whisker range in each cluster, and diamonds are stacks in two clusters at high redshifts. The lower panel shows the differences between various EPS models and {\Yehb}. {\it Right:} Redshift evolution of the rest-frame FUV-$r$ colors (for various EPS models) and the observed NUV-$i$ colors (for observational data points). The lines are the same as in the left panel. Data points are the observational results of field red galaxies of \protect\cite{DePropris22} and \protect\cite{Atlee09}, and their error bars illustrate the root mean squares of the mean values from 100 bootstrapping realizations within each magnitude/redshift bin.}
  \label{fig:redshift}
\end{figure*}

\subsection{Colors and redshift evolution of model galaxies}
\label{section:colors}

\autoref{fig:SAM} displays the distributions of all model galaxies at $z\sim0$ on the FUV-NUV versus NUV-$r$ color diagram, obtained by adopting different EPS models as indicated. Black contours show galaxy number density, while filled colors depict the mean specific star formation rate (sSFR)\footnote{For modeled galaxies with sSFR lower than $10^{-13}\peryr$, we set their sSFR to $10^{-13}\peryr$. In observations, the uncertainty in measuring sSFR for galaxies with sSFR less than $10^{-13}\peryr$ is large and not well determined \citep[see Fig. 15 in][]{Salim07}. Thus, in our semi-analytic model, although sSFR can be estimated down to $10^{-20}\peryr$, we set $10^{-13}\peryr$ as the lower limit for comparability with observations. In reality, these galaxies represent exhausted systems with almost no star formation.} within each cell. For {\YunnanII} and {\sc galaxev}, the fiducial or default IMF is applied, while for {\sc bpass}, we use the IMF with a slope of $\Gamma=1.35$ for both single and binary population models.
For comparison, we selected a volume-limited sample of elliptical galaxies with both optical and UV photometry from the NASA-Sloan Atlas \citep[NSA,][]{NSA}, a local galaxy catalogue constructed from the Sloan Digital Sky Survey \citep[SDSS,][]{SDSS} and {\sls Galaxy Evolution Explorer} \citep[{\sls GALEX},][]{GALEX}. Our sample comprises 35,336 elliptical galaxies with redshift $0.01 < z < 0.05$, stellar mass $\Mstar > 10^{9.5} \Msun$, and Sersic index $n_{\rm s} > 2.5$. The distribution of this sample is plotted as purple contours, repeated in every panel of~\autoref{fig:SAM}. In each panel, dashed lines indicate the empirical criteria suggested by \cite{Yi11} to identify UV-upturn galaxies, which are located in the lower-right corner of the diagram, with NUV-$r > 5.4$, FUV-NUV $< 0.9$, and FUV-$r<6.6$.

As can be seen, all models exhibit a bimodal distribution in both color indices, with a clear division between red and blue populations at NUV-$r\sim5$, similar to the sample of real galaxies. All models span a similar range in NUV-$r$, comparable to real galaxies, but the FUV-NUV colors in both red and blue populations are limited to narrower ranges than in the real sample. Consequently, the models lack galaxies with extreme colors, e.g., those with FUV-NUV$\gtrsim 4$ or $\lesssim -0.5$. Evidently, none of the models can fully reproduce the distribution of real galaxies in this diagram, indicating the presence of systematic uncertainties within the models. This inconsistency is not unexpected, considering that galaxy colors are not included in the model calibration. We will revisit and discuss this issue in detail in~\autoref{section:discussion-limits}. In what follows, we will focus on the relative changes in UV and optical colors caused by various physical processes.

\begin{figure*}
  \centering
  \includegraphics[width=1.0\textwidth]{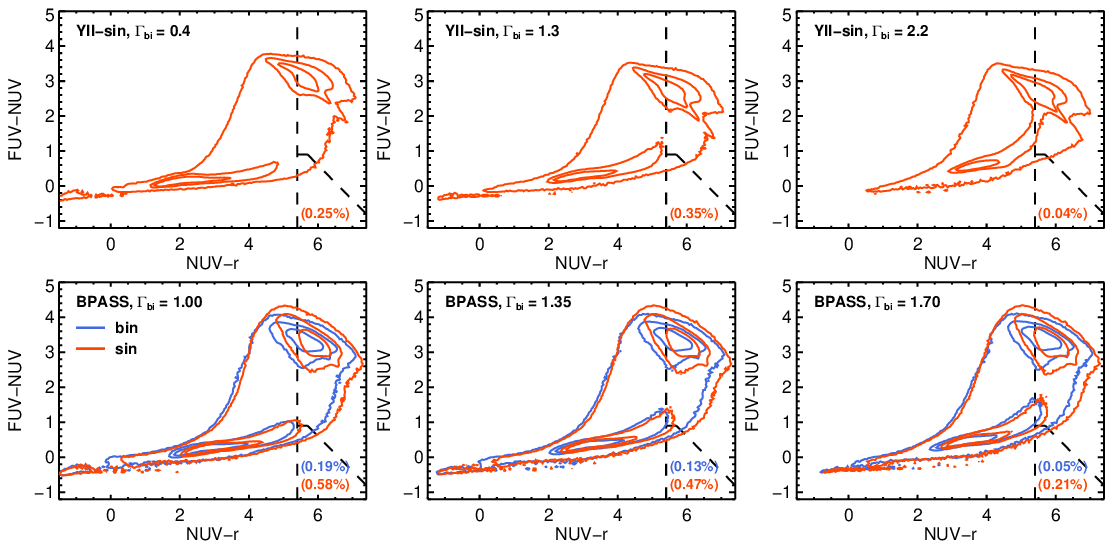}
  \caption{Distributions of modeled galaxies run with {\Ysin} (upper panels) and {\sc bpass} models (lower panels) with different IMF high-mass end slopes $\Gamma_{\rm bi}$, as indicated by different contours. Percentages in brackets are the number fractions of UV-upturn galaxies over all the modeled galaxies.}
  \label{fig:IMF}
\end{figure*}

The major difference among the models lies in the FUV-NUV color of red populations. The {\sc bpass} models of both single and binary populations predict redder FUV-NUV colors than the {\YunnanII} models. Among other models, the {\Yehb} model predicts the bluest FUV-NUV colors for the red population. As a result, a tail of the red population in the {\Yehb} model falls into the UV-upturn region, although the fraction is much smaller than that of UV-upturn galaxies in the real sample due to the aforementioned lack of extreme colors in the models. The fractions of UV-upturn galaxies in the other models are significantly smaller than in the {\Yehb} case. This echoes the theoretical expectation that the formation of EHB stars through binary interactions can indeed be an efficient way to produce UV excess in elliptical galaxies. Interestingly, in both {\Ybin} and {\sc bpass}, the single star model and binary model behave similarly in the color-color diagram, with only slight differences in NUV-$r$ and almost no difference in FUV-NUV for both red and blue populations.

We now extend the analysis to higher redshifts. The left panel of~\autoref{fig:redshift} shows the redshift evolution of the rest-frame NUV-$r$ color for quenched galaxies (sSFR $<10^{-12}\peryr$) in different EPS models. The {\Yehb} and {\sls BPASS-sin} models yield the bluest and reddest NUV-$r$ colors, respectively; therefore, these two models are fully illustrated as thick solid lines with corresponding distributions (shaded regions: $25\%-75\%$ quartiles; dashed lines: an additional 1.5 times the interquartile range) to indicate the plausible range predicted by different EPS models. The median values of other EPS models are marked as dotted lines for clarity. For comparison, box plots show the distribution of ETGs within 12 galaxy clusters at $0.3<z<1.0$ from the Hubble Legacy Archive \citep{Ali21}, whose boxes and whiskers share the same definition as our shaded regions and dashed lines, respectively. Circles are observed outliers beyond the whisker range in each cluster, and diamonds are stacks in two clusters at high redshifts. While this study does not aim to quantitatively compare our models with observations, it is noteworthy that the model derived with {\Yehb} agrees quite well with the observational results, in terms of both the median and the scatter of NUV-$r$ at a given redshift. However, it is important to note that the differences across different EPS models are comparable to or even smaller than the inherent scatter of galaxies.

In the right panel of~\autoref{fig:redshift}, we compare the predicted redshift evolution of the "observed" NUV-$i$ color of quenched model galaxies with field red galaxies from \cite{DePropris22}. Following \cite{DePropris22}, we use the rest-frame FUV-$r$ color from the EPS models as analogues of the observed NUV-$i$ colors given their redshift ranges\footnote{We find the measured NUV-$i$ colors are in fact $\sim1-2$ magnitudes bluer than the rest-frame FUV-$r$ colors due to the broader bandwidth of the NUV band \citep{Morrissey05}. We have corrected this effect for each modeled galaxy according to its redshift and mean mass-weighted age and metallicity, by utilizing spectra of each EPS model.}. As observed, the {\Yehb} model exhibits marginal agreement with the upper limits of the observational results, whereas other EPS models yield excessively redder colors. The variations of the rest-frame FUV-$r$ color among the EPS models are significantly larger than those of the rest-frame NUV-$r$ color seen in the left panel, especially at high redshift. This is consistent with what we have seen from~\autoref{fig:SSP_solar}, where the FUV-involved colors are more powerful than NUV-involved colors in distinguishing different EPS models at old ages. \cite{Ali21} and \cite{DePropris22} suggested that the He-rich single star model provided the best explanation for their observations. Our analysis further suggests that the binary star model {\Yehb} is also compatible with current observations in terms of the redshift evolution of quenched galaxies at $z\lesssim1$.

\begin{figure*}
\centering
\includegraphics[width=0.95\textwidth]{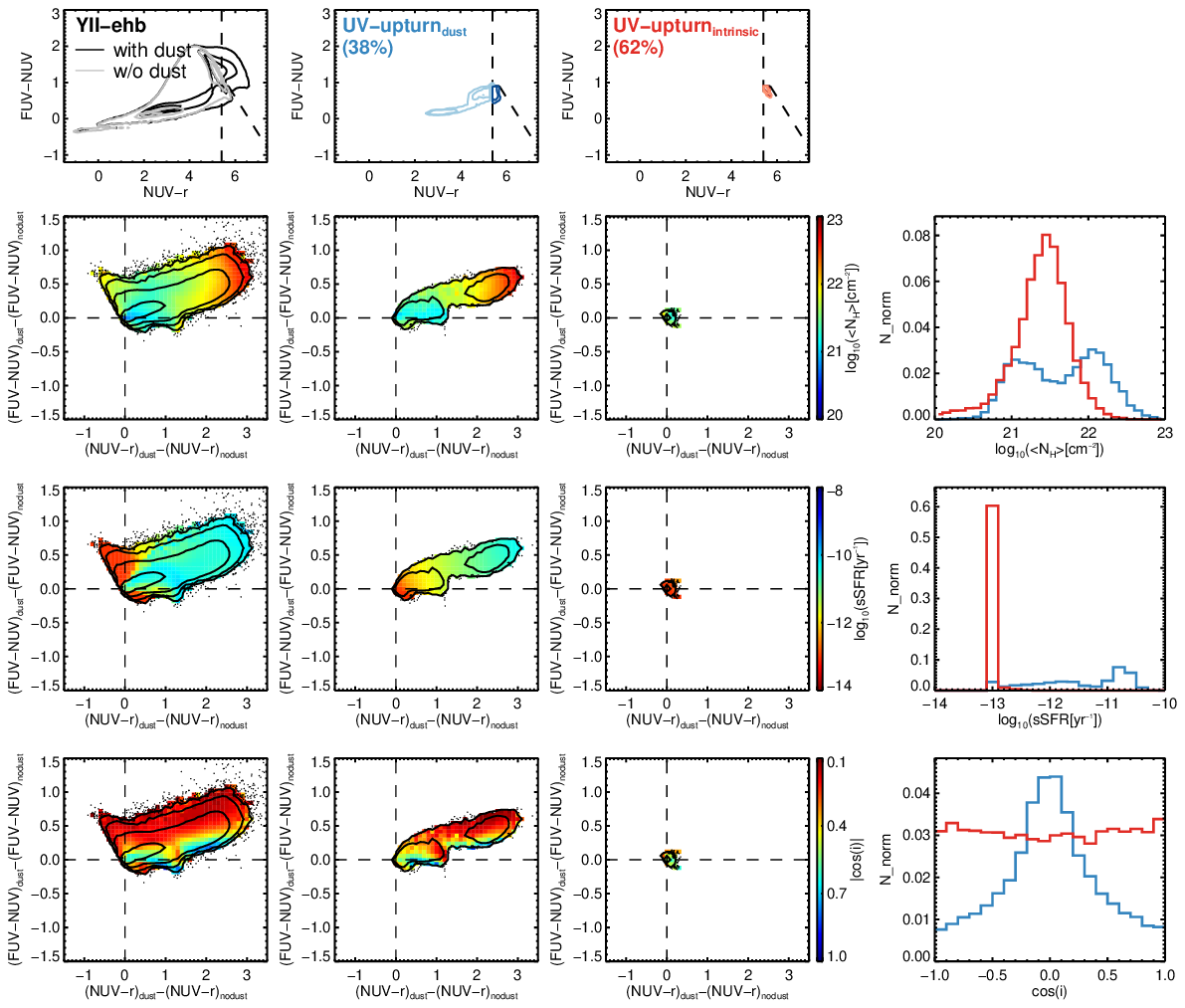}
\caption{Top panels: Distributions of all modeled galaxies (black), {\UVXdust} galaxies (blue) and {\UVXintrinsic} galaxies (red) on the color-color diagram, by using {\Yehb} model. Dark/light contours show the ones with/without dust attenuation. Lower panels: Distributions of corresponding reddenings for each kind of galaxies. Black contours show number ratios, and filled colors represent mean H column density $\left<N_{\text H}\right>$, sSFR, and inclination angle $|\cos(i)|$ respectively as indicated by color bars. Distributions of these galactic properties are shown in the right column, normalized by the total number of two kinds of UV-upturn galaxies.}
\label{fig:dust_UV}
\end{figure*}

\subsection{Influence of stellar initial mass function}
\label{section:IMF}

The intensity of the UV-upturn phenomenon should, in principle, depend on the high-mass end of the adopted stellar initial mass function (IMF), which determines the fraction of stars more massive than $\sim0.6\Msun$, including the progenitors of EHB stars ($0.8\lesssim\mpri\lesssim5\Msun$, \citealt{Han02,Clausen12}). To investigate this effect, we adopt the "bimodal" IMF form \citep{Vazdekis96} to describe IMFs with different slopes at the high-mass end:
\beq
\Phi(\log_{\rm 10}{(\mpri)}) \propto
\begin{cases}
(\mpri/0.6)^{-\Gamma_{\rm bi}}& \text{(for $\mpri>0.6\Msun$)}\\
p(\mpri)& \text{(for $0.2 \le \mpri \le 0.6\Msun$),}\\
1& \text{(for $\mpri<0.2\Msun$)}
\label{equ:IMF}
\end{cases}
\eeq
where $\mpri$ is the primary mass in unit of $\Msun$, $\Gamma_{\rm bi}$ is the slope for $\mpri>0.6\Msun$, and $p(\mpri)$ is a third-degree spline function connecting the low- and high-mass ends. The slope $\Gamma_{\rm bi}$ is varied within the range $[0.4,2.2]$ to mostly encompass the slopes estimated by \cite{Zhou19} for a sample of elliptical galaxies from the MaNGA survey. All IMFs are assumed to be truncated below $0.1\Msun$ and above $100\Msun$, and normalized within the range $[0.1,100]\Msun$. With this form, the bimodal IMF of $\Gamma_{\rm bi} = 1.3$ is close to the fiducial IMF adopted in {\GABE} \citep[][hereafter MS79]{MS79} at masses up to $\sim10\Msun$ and the commonly-used IMF of \cite{Chabrier03} at all masses.
We have also repeated our analysis using IMFs other than the bimodal IMFs, as well as single power-law IMFs ($\Phi(\log_{\rm 10}{(\mpri)}) \propto \mpri^{-\Gamma_{\rm si}}$), including the IMF of \cite{Salpeter55}. We find that the specific form of IMFs does not influence our conclusion. Therefore, we only present results from the bimodal IMFs for clarity. A uniform mass-ratio distribution of the secondary and the primary mass ($\msec/\mpri$) is assumed \citep{Mazeh92,Goldberg94}, as the same in the fiducial {\YunnanII} models.

The implementation of various IMFs for {\Ybin} and {\Yehb} is not trivial, and the resulting magnitudes are extremely sensitive to model parameters \citep[e.g.][]{GR90,Dorman95,Han02}. Thus, for {\YunnanII}, we explore the effect of IMF variation only for the models of {\Ysin}.  In {\sc bpass} v2.2, a variety of broken power-law IMFs are available for both single and binary populations \citep[][see their Tab. 1]{Stanway18}, which share a similar form with~\autoref{equ:IMF} but with marginally steeper slopes at the low-mass end. To be consistent with {\YunnanII} models, we use {\sc bpass} data with upper mass cut-offs of $100\Msun$. In this way, we have $\Gamma_{\rm bi}=1.0, 1.35, 1.7$ for both single and binary models from {\sc bpass}, as well as a range of $0.4\leq \Gamma_{\rm bi}\leq2.2$ for single models from {\Ysin}. 
With each of the various IMFs and the corresponding EPS models, we run the semi-analytic model code {\GABE}. The resulting galaxy samples are plotted on the FUV-NUV versus NUV-$r$ diagram in~\autoref{fig:IMF}.

As $\Gamma_{\rm bi}$ increases, galaxies within the star-forming sequence become redder in both NUV-$r$ and FUV-NUV color indices, while red population galaxies exhibit minimal change, becoming only slightly bluer in FUV-NUV. This pattern holds true for both the {\Ysin} model and the single model of {\sc bpass}. The binary model in {\sc bpass} similarly follows this trend with $\Gamma_{\rm bi}$ as seen in the single model. The weak dependence of the red population on the initial mass function (IMF) can be attributed to the dominance of long-lived, low-mass stars in these galaxies, which are not substantially affected by changes in the high-mass end slope of the IMF.

When examining the impact on the formation efficiency of UV-upturn galaxies, different behaviors emerge across various evolutionary population synthesis (EPS) models. In the {\Ysin} model, the configuration with $\Gamma_{\rm bi}=1.3$ exhibits the highest efficiency. Conversely, in both the single and binary {\sc bpass} models, top-heavy IMFs (those with lower $\Gamma_{\rm bi}$) display higher UV-upturn fractions, with $\Gamma_{\rm bi}=1.00$ being the most efficient. These inconsistencies suggest systematic uncertainties related to IMF variations among different EPS models. However, the differences in UV-upturn fractions remain within an order of magnitude. Compared to the influence of extreme horizontal branch (EHB) stars, variations in the IMF appear to have a secondary effect on the formation of UV-upturn galaxies. This suggests that a universal IMF should suffice for investigating the UV-upturn phenomenon.

The redshift evolution of the rest-frame NUV-$r$ and FUV-$r$ colors of modeled quenched galaxies are presented in Appendix~\ref{appendix:redshift_IMF}, for EPS models with different IMF slopes. Consistent with the result obtained above, variation in IMF slope has a negligible effect on the UV-to-optical colors in old stellar populations.

\subsection{Influence of dust attenuation}
\label{section:dust}

In our analysis thus far, we have not discussed the effects of dust attenuation, which can absorb and scatter UV and optical light, significantly altering galaxy colors and potentially leading to the incorrect identification of UV-upturn galaxies. To incorporate dust attenuation, {\GABE} uses the model proposed by \cite{DeLucia07}, which considers the reddening effects of both the diffuse interstellar medium (ISM) and birth clouds around young stars. For the ISM, reddening is modeled with a "slab" geometry as described by \cite{Devriendt99}, while the reddening of birth clouds follows the model from \cite{Charlot00} and \cite{DeLucia07}. For more comprehensive details on the dust attenuation model, readers are referred to \cite{DeLucia07} and the references therein. Essentially, the attenuation at a given wavelength for both the ISM and birth clouds is related to the mean optical depth for a face-on average disk, $\tau_{\lambda}$, which depends on the chosen dust attenuation curve ($A_{\lambda}$), gas metallicity ($Z_{\text{g}}$), and mean hydrogen column density ($\left<N_{\text H}\right>$). Additionally, ISM reddening depends on the dust albedo (\citealt{Mathis83}) and the inclination angle of the disk ($\cos(i)$), defined as the angle between the direction of the spin of the gaseous disk and the line of sight, assumed to be along the third axis of the N-body simulation.

\begin{table}
  \caption{Parameters of dust attenuation curves. The parameter $B \equiv A_{\rm bump}/A_{\rm 2175}$ is the fraction of UV bump over the total extinction at $2175$ Å, as defined in \protect\cite{Salim20}.}
  \begin{tabular}{lccc}
    \hline
    Model & slope $\delta$ & bump amplitude $\Eb$ & Description\\
    \hline
    M83 & - & - & \cite{Mathis83}, fiducial \\
    \hline
    E130 & $0.0$ & $13.0$ & corresponding to $B=0.6$\\
    E035 & $0.0$ & $3.5$ & $B=0.3$, Milky Way/LMC\\
    E000 & $0.0$ & $0.0$ & $B=0.0$, Calzetti00\\
    \hline
    Dm06 & $-0.6$ & $1.99$ & \multirow{2}{*}{$\Eb = -1.9\delta + 0.85$}\\
    D00 & $0.0$ & $0.85$ & \multirow{2}{*}{\citep{Kriek13}} \\
    Dp03 & $+0.3$ & $0.28$ & \\
    \hline
    \label{tab:dust_models}
  \end{tabular}
\end{table}

The impact of dust attenuation is illustrated in the top-left panel of~\autoref{fig:dust_UV}, which compares the color-color diagrams of model galaxies with and without dust attenuation. The lower panels in the leftmost column show the distribution of changes in the two color indices due to dust attenuation, color-coded by $\left<N_{\text H}\right>$, sSFR, and $|\cos(i)|$. For simplicity and clarity, we present only the results for the {\Yehb} model, noting that other EPS models yield similar outcomes. As demonstrated, dust attenuation generally causes model galaxies to appear redder in both FUV-NUV and NUV-$r$ colors, with greater reddening associated with higher values of $\left<N_{\text H}\right>$, sSFR, and inclination angle. To assess the importance of dust attenuation in identifying UV-upturn galaxies, we divide all UV-upturn galaxies into two groups: {\UVXintrinsic}, which are intrinsic UV-upturn galaxies that remain in the UV-upturn region after accounting for dust attenuation, and {\UVXdust}, which includes galaxies that enter the UV-upturn region only after accounting for dust attenuation.

The second and third columns of~\autoref{fig:dust_UV} show color-color diagrams and color change distributions for the two  categories, while the rightmost column displays histograms of the parameters used to color-code the distributions of color changes. The {\UVXintrinsic} group contains 62\% of the UV-upturn galaxies, primarily quenched galaxies with minimal star formation and intermediate $\left<N_{\text H}\right>$. Galaxies in the {\UVXdust} group which consists of 38\% of the UV-upturn sample are mainly star-forming, spanning a wide range of sSFR, exhibiting a bimodal distribution in $\left<N_{\text H}\right>$, and a narrow inclination angle range centered at $i=90^{\circ}$. The subset with higher $\left<N_{\text H}\right>$ and sSFR experiences more significant reddening, thus shifting from the star-forming sequence into the UV-upturn region. The other subset, nearly quenched with low sSFR, also enters the UV-upturn region, although their reddening is moderate. In addition, we notice that approximately one-fifth of intrinsic UV-upturn galaxies relocate outside the UV-upturn region when dust attenuation is taken into account. Most of these are quenched disk galaxies. Unlike their elliptical counterparts, these quenched disk galaxies are more susceptible to the effects of dust attenuation and may exit the UV-upturn region if their FUV-NUV color becomes excessively red.

\begin{figure*}
\centering
\includegraphics[width=0.8\textwidth]{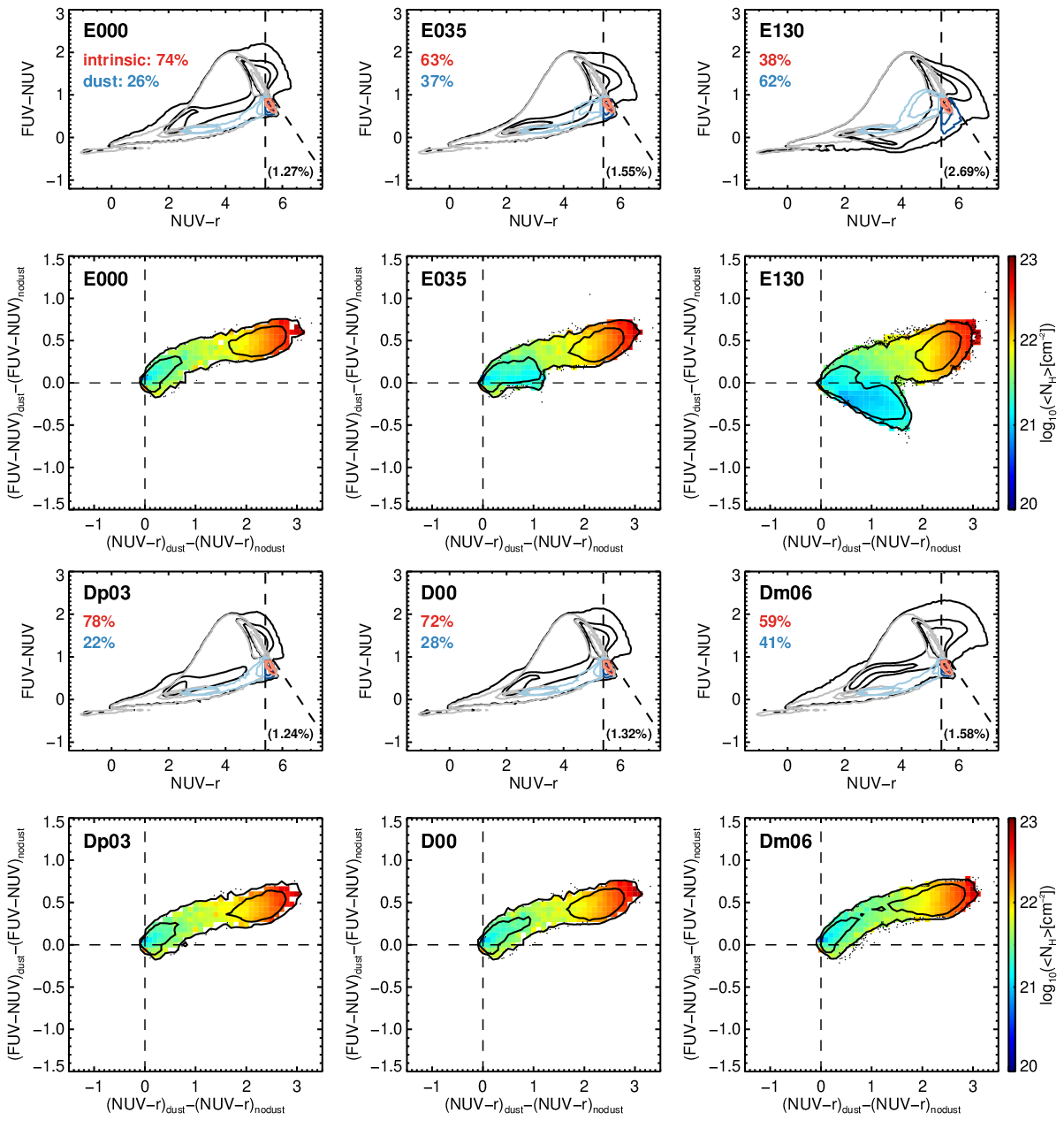}
\caption{The same as the top row and second column in~\autoref{fig:dust_UV}, but for modeled galaxies with different dust attenuation curves as parameterized in~\autoref{tab:dust_models}.}
\label{fig:dust_bump}
\end{figure*}

By default {\GABE} adopts the attenuation curve of \citet[][M83 hereafter]{Mathis83}, an average observational relation of stars in the Milky Way \citep{Savage79}, thus exhibiting a noticeable bump at 2175\AA\ and a relatively steep slope in UV, typical of attenuation curves in Milky Way \citep{Salim20}. However, previous studies of local galaxies have revealed a wide range of UV slope and UV bump strength \citep[see][for a review]{Salim20}. The FUV and NUV magnitudes are expected to be modified differently by attenuation curves of different slopes and UV bumps. To evaluate this effect, we adopt the parameterized formula proposed by \cite{Noll09} to describe various dust attenuation curves:
\beq
A_{\rm \lambda} = \frac{A_{\rm V}}{4.05} \left( k^{\prime}_{\rm \lambda}+D_{\rm \lambda}\right) \left( \frac{\lambda}{\lambda_{\rm V}} \right)^{\delta},
\label{eq:Alambda}
\eeq
where $k^{\prime}_{\rm \lambda}$ is the bumpless attenuation curve obtained from local starburst galaxies by \cite{Calzetti00}, hereafter Calzetti00, and $D_{\rm \lambda}$ is a Lorentzian-like Drude profile to parameterize the UV bump, given by
\beq
D_{\rm \lambda} = \frac{E_{\rm b} \left( \lambda\Delta\lambda\right)^2}{\left( \lambda^2-\lambda_{\rm 0}^2 \right)^2 + \left( \lambda\Delta\lambda \right)^2}.
\label{eq:Dlambda}
\eeq
There are four parameters in this parameterization: the slope $\delta$ accounts for additional wavelength dependence that diverges from the Calzetti00 curve; $E_{\rm b}$, $\lambda_{\rm 0}$ and $\Delta\lambda$ denote the amplitude, central wavelength, and full width at half maximum (FWHM) of the UV bump, respectively \citep{Fitzpatrick90}. The latter two parameters are set to average values, specifically $\lambda_{\rm 0}=2175$ Å and $\Delta\lambda=350$ Å \citep{Noll09b,Seaton79,Kriek13}, and we investigate the parameter space of the slope $\delta$ and the UV bump amplitude $E_{\rm b}$. We consider two sets of models of dust attenuation curves with varying parameters $\delta$ and $\Eb$ as listed in~\autoref{tab:dust_models}. The first set is aimed to isolate the effects of UV bump strength, with a fixed slope of $\delta=0$ but three different values of $\Eb$. The second set of models are assumed to follow the observed relation of $\Eb = -1.9\delta + 0.85$ from \cite{Kriek13}. Graphs of these curves are presented in Appendix~\ref{appendix:redshift_dust}.

\autoref{fig:dust_bump} presents the color-color diagram and illustrates how model galaxies' colors change with different dust attenuation curves. For each model, the upper panel shows the color-color diagram, with contours in black, red, and blue representing all model galaxies and specifically those in the {\UVXintrinsic} (red) and {\UVXdust} (blue) groups, while the lower panel highlights the shifts in two color indices of {\UVXdust} galaxies, with changes color-coded according to $\left<N_{\text H}\right>$. The percentages in black brackets indicate the ratio of UV-upturn galaxies to the total modeled galaxies, whereas the red and blue percentages denote the proportions of {\UVXintrinsic} and {\UVXdust} galaxies within the UV-upturn category, respectively. When comparing the E000, E035, and E130 models, we observe that the proportion of UV-upturn galaxies rises as the UV bump's strength increases, a trend primarily driven by {\UVXdust} galaxies. This effect is most evident in the E130 model, where the influence of the UV bump leads some green valley galaxies to exhibit bluer FUV-NUV and redder NUV-$r$ colors. Consequently, these galaxies shift towards the lower-right corner of the color-color diagram along the quenched sequence, eventually reaching the UV-upturn region. In contrasting the Dp03, D00, and Dm06 models, where UV bumps are relatively weak, we find that steeper attenuation curves result in greater reddening of both NUV-$r$ and FUV-NUV colors. This alteration makes it easier for star-forming galaxies to transition into the UV-upturn region.

The redshift evolution of the UV-to-optical colors of quenched galaxies with various dust attenuation models are discussed in Appendix~\ref{appendix:redshift_dust}. We find that different dust models merely have a negligible influence, with $\Delta$(NUV-$r$) $\lesssim0.10$ and $\Delta$(FUV-$r$) $\lesssim0.15$, as these quenched galaxies are naturally devoid of dust.

While elliptical galaxies were usually thought to be dust-free, recent studies suggest that residual star formation (RSF) and associated dust attenuation are not uncommon within these galaxies. \cite{Kaviraj07} found that approximately 30\% of ETGs exhibit UV-to-optical colors indicative of some degree of RSF activity occurring within the last Gyr. Additionally, \cite{Vazdekis16} noted that young stellar populations with ages between 0.1 and 0.5 Gyr contribute mass fractions of 0.1\% to 0.5\% to massive ETGs, based on their analysis of UV colors and line strengths. Similarly, \cite{Salvador20} reported that RSF is prevalent in massive ETGs, with an average mass fraction of around 0.5\% in young stars formed within the last 2 Gyr. Notably, \cite{Werle20} detected young stellar components in 17.5\% of the UV-upturn galaxies defined by \cite{Yi11}, which aligns closely with our lower limit in this study. A more in-depth discussion of the role of dust attenuation in the context of UV-upturn galaxies can be found in~\autoref{section:discussion-dust}.

\subsection{Joint dependence on $\Fb$, $\Fy$, $\Zstar$ and dust}
\label{section:joint}

\begin{figure}
  \centering
  \includegraphics[width=0.48\textwidth]{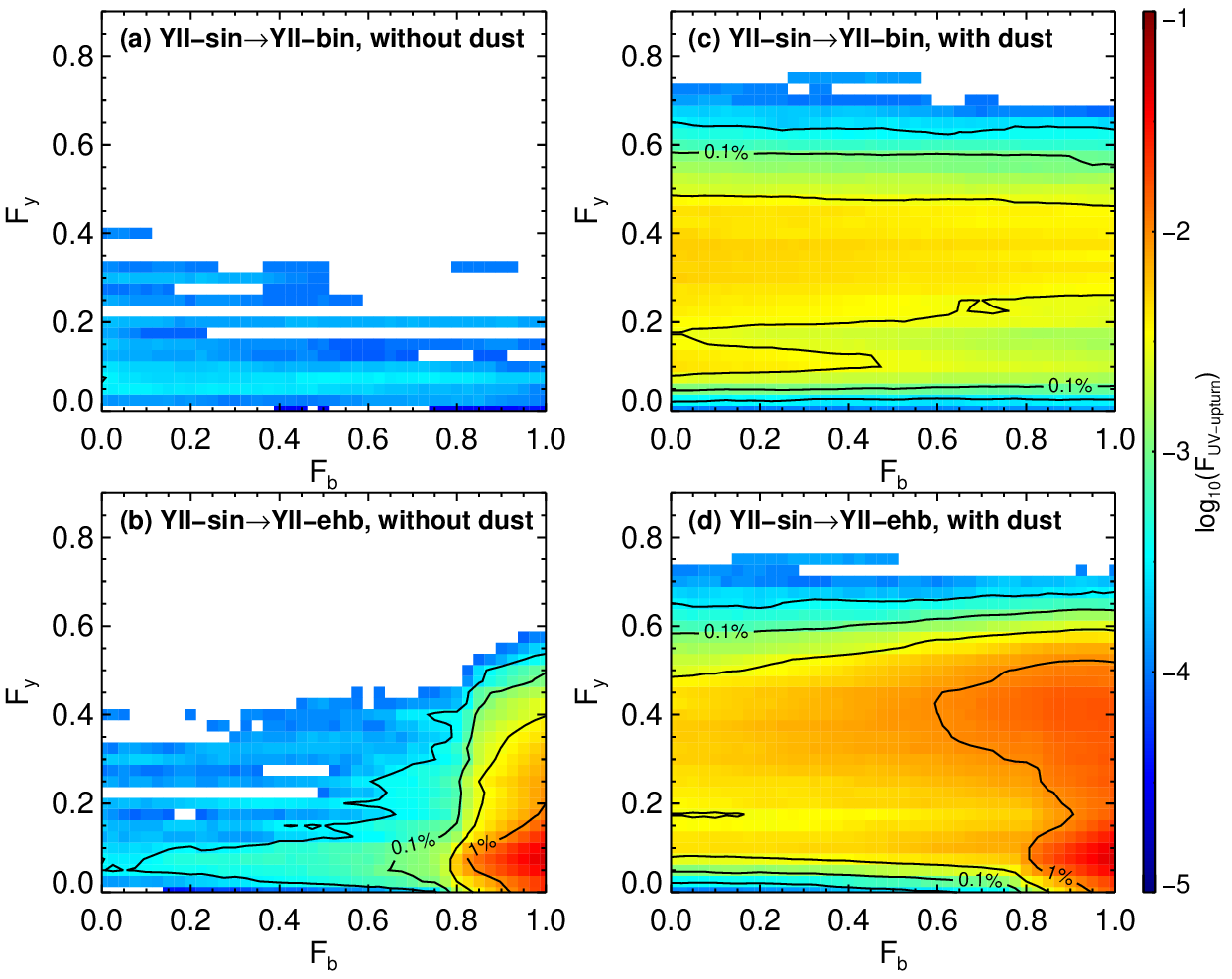}
  \caption{Filled colors in each cell indicate fractions of UV-upturn galaxies ($F_{\rm UV-upturn}$) as a function of binary mass fraction ($\Fb$) and young fraction ($\Fy$). The latter one is defined as the mass fraction of SSPs which have mass-weighted age younger than $9\Gyr$. Cell size is $0.025\times0.025$, and only cells with more than 5 galaxies are presented. Contours correspond to the $0.03\%$, $0.1\%$, $0.3\%$ and $1\%$ of $F_{\rm UV-upturn}$. Binary interactions without the formation of EHB stars are shown in the upper panels, and ones with EHB stars are in the bottom panels. Results without dust attenuation are in the left column, and ones with dust are in the right column.}
  \label{fig:Fb_Fy}
\end{figure}

In the previous subsections, we explored how UV-upturn galaxies are influenced by stellar evolution, which is described through the photometric evolution of SSPs derived from EPS models, as well as by dust attenuation. These are the two primary physical processes involved in calculating magnitudes for galaxies in semi-analytical models. We will now focus on the fiducial {\YunnanII} models and the M83 dust attenuation curve to further examine the dependence on various galaxy properties, including the fraction of binary populations ($\Fb$), the fraction of young populations ($\Fy$), and stellar metallicity ($\Zstar$), considering both scenarios with and without dust attenuation. Here, $\Fb$ is defined as the mass fraction of binary populations relative to the total stellar population, while $\Fy$ refers to the mass fraction of SSPs with a mass-weighted age younger than 9 Gyr. It is important to note that using an alternative criterion for $\Fy$ only affects the detailed scale of the figures and does not change our overall conclusions.

\autoref{fig:Fb_Fy} illustrates the distributions of $F_{\rm UV-upturn}$, defined as the proportion of UV-upturn galaxies corresponding to specific values of $\Fb$ and $\Fy$, plotted on the $\Fy$ versus $\Fb$ plane. The upper panels show models of binary interactions that exclude the formation of extreme horizontal branch (EHB) stars. In these models, varying $\Fb$ from 0 to 1 transitions from the {\Ysin} model to the {\Ybin} model by progressively incorporating a $\Fb$ fraction of binary populations. In contrast, the lower panels present models that include EHB stars, transitioning from the {\Ysin} model to the {\Yehb} model. Results without dust attenuation are shown in the left column, where we find that a significant population of UV-upturn galaxies can form only when including binary interactions with EHB stars within old SSPs. Results incorporating dust attenuation are displayed in the right column, revealing a notably larger fraction of UV-upturn galaxies compared to the left panels. Similar to the no-dust scenario, EHB stars contribute to the peak observed around $\Fb > 0.8$ and $\Fy \sim 0.1$, representing the UV-upturn phenomenon in nearly fully quenched galaxies. Additionally, dust attenuation creates a stripe corresponding to a roughly constant young fraction of $\Fy \sim 0.4$, which signifies star-forming galaxies. The influence of EHB stars within old SSPs is evident in both the region representing star-forming galaxies and that of quenched galaxies.

\begin{figure}
  \centering
  \includegraphics[width=0.48\textwidth]{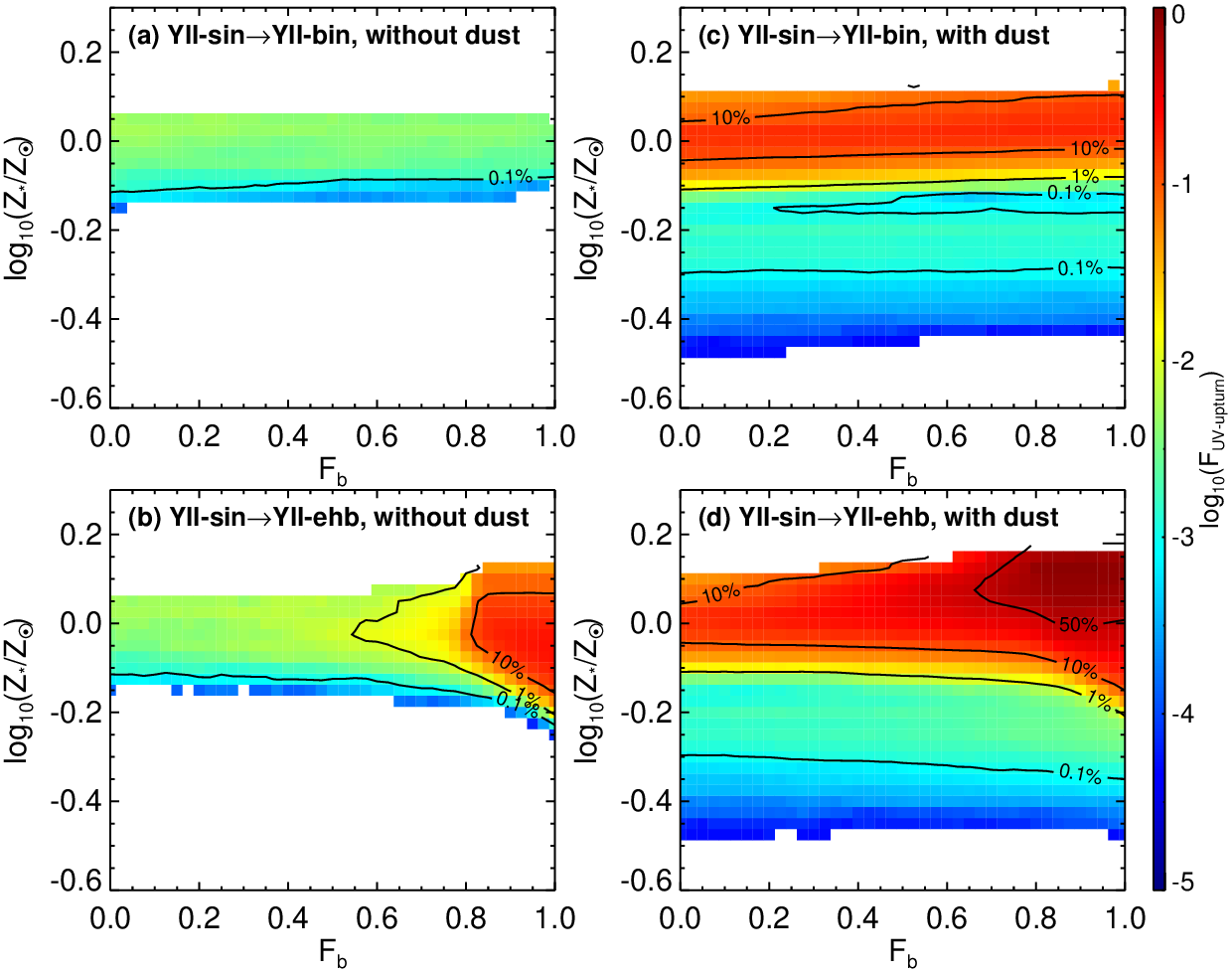}
  \caption{The same as in Fig.~\ref{fig:Fb_Fy}, but as a function of binary mass fraction ($\Fb$) and stellar metallicity ($\Zstar$). Contours correspond to the $0.1\%$, $1\%$, $10\%$ and $50\%$ of $F_{\rm UV-upturn}$.}
  \label{fig:Fb_metal}
\end{figure}

\autoref{fig:Fb_metal} further explores the joint dependence on $\Fb$ and stellar metallicity $\Zstar$. Notably, there is a strong preference for high stellar metallicity in both categories of UV-upturn galaxies: those formed through binary interactions with EHB stars and those resulting from dust attenuation. This finding is consistent with our expectations derived from analyzing SSPs in the {\Yehb} model. As discussed in~\autoref{section:EPS}, only the most metal-rich SSPs with $Z > 0.01$ can evolve into the UV-upturn region after approximately 4 Gyr. In contrast, the latter preference emerges from the complex interactions of star formation activities integrated into semi-analytical and dust attenuation models. In the most extreme case, over 80\% of galaxies with $\Fb > 0.7$ and $\log_{10}{(\Zstar/\Zsun)} \sim 0.1$ in panel (d) are classified as UV-upturn galaxies, despite the relatively small absolute number of such metal-rich galaxies. This strong preference for high stellar metallicity accounts for the observed correlation between the UV-upturn phenomenon and metallicities \citep[e.g.][]{Carter11, Smith12, Akhil24}. Specifically, we find that the mean stellar metallicities of UV-weak and RSF galaxies in the models are approximately 0.15 to 0.2 dex lower than those of UV-upturn galaxies. This aligns well with the observed metallicity differences between UV-upturn and UV-weak regions, as reported in \cite{Jeong12}. We will further discuss on the implications of the metallicity dependence of the UV-upturn phenomenon in~\autoref{section:discussion-binary}.

\section{Discussion}
\label{section:discussion}

\subsection{Influence of the definition of ``UV-upturn''}
\label{sec:discussion_definiton}

The primary prediction of the "binary star model" \citep{Han07} was that the UV-upturn phenomenon exhibits only a weak dependence on the age and metallicity of stellar populations. However, upon revisiting the binary star model within a cosmological context, we find that this prediction requires more nuanced consideration. In this work, we find that binary star model exhibits a strong dependence on both the age and metallicity of stellar population. This discrepancy with \cite{Han07} is actually mostly attributed to our different definition of ``UV-upturn''.

As pointed out in \autoref{section:definition}, the definition or measurement of the UV-upturn phenomenon varies from one paper to another. The relatively stringent photometric definition of UV-upturn galaxies of \cite{Yi11} has been adopted in some observational \citep[e.g.][]{Jeong12,Petty13,Boissier18} and theoretical \citep[e.g.][]{HB14} works to physically select UV-upturn galaxies. While in \cite{Han07} and many other works, the value of the FUV-$r$ color (or other UV-to-optical color) alone was used to quantify the strength of UV excess \citep[e.g.][]{Brown03,Donahue10,Smith12,Phillipps20,Ali21}; likewise, in some other works, the FUV-NUV color alone was used \citep[e.g.][]{Loubser11,Carter11}. These choices were made sometimes due to the lack of photometric data at certain wavelengths, or because the sample volume might be largely reduced with additional color criteria. Moreover, observational systematic uncertainties further constrain the power of rather stringent criteria. For example, when the UV-upturn criteria of \cite{Yi11} were applied to the Coma cluster, \cite{Yi11} found that none of the 30 brightest elliptical galaxies falls into the UV-upturn region. In contrast, \cite{Smith12} reported that 14 out of 150 quenched galaxies in the Coma cluster meet the criteria, and this discrepancy was likely due to their different applications of k-correction. Furthermore, in some spectroscopic studies of ETGs at high redshift \citep[e.g.][]{Lonoce20}, line indices were used in SED fitting to ascertain whether a galaxy favors an old stellar population with an upturn contribution, in which case it would be defined as a UV-upturn galaxy. \cite{LeCras16} found that the fraction of UV-upturn galaxies defined in this way was approximately seven times that reported in \cite{Yi11}, as some UV-weak galaxies might also select an upturn SSP if only the line indices' strength were considered.

Here, we emphasize that the definition or measurement of the UV upturn is crucial when addressing the so-called ``UV excess'' phenomenon. Theoretical models can exhibit significantly different dependencies on age and metallicity based on the definitions employed. For instance, using the triple color criteria established by \cite{Yi11}, the UV-upturn phenomenon associated with the binary star model shows a strong dependence on both age and stellar metallicity. In contrast, when employing the FUV-$r$ color as a measure of the UV upturn's strength, as done by \cite{Han07}, the dependence on age and metallicity appears relatively weak, as illustrated in the right panel of~\autoref{fig:SSP_solar} and middle panel of~\autoref{fig:SSP_metal}. Conversely, utilizing the FUV-NUV color again suggests a strong dependence on both age and metallicity. Thus, it is essential to clearly define the term ``UV-upturn'' prior to any discussion, and to ensure consensus on this definition when comparing observations with theoretical models.

\subsection{Metallicity dependence}
\label{section:discussion-binary}

\cite{Han07} concluded that the formation of extreme horizontal branch (EHB) stars is primarily determined by their initial orbital parameters, with metallicity playing a minimal role in the related mass transfer processes. As a result, the UV-upturn phenomenon was expected to show weak or no correlation with metallicity. However, they also pointed out that metallicity ``may affect the properties of the binary population in more subtle ways''. Actually, the total flux of a stellar population consists of contributions not only from EHB stars but also from all other stars. In the {\Yehb} model, \cite{Zhang10} incorporated the binary formation channels of EHB stars into the {\YunnanII} model, under the assumption that the formation efficiencies of EHB stars remain unaffected by metallicity. This implies that the absolute contributions of EHB stars are consistent across SSPs with varying metallicities in the {\Yehb} framework. As illustrated in the middle panels of~\autoref{fig:SSP_metal}, with this straightforward assumption, UV-upturn phenomenon resulting from binary star model actually exhibits a dependence on metallicity. Adopting the stringent photometric definition of \cite{Yi11}, the dependence is quite strong. Only metal-rich SSPs (with metallicity greater than $0.01 \Zsun$) can reach the UV-upturn region as they age beyond approximately 4 Gyr. In more metal-poor SSPs, the NUV flux from stars other than EHB stars becomes excessively dominant, thereby impeding the fulfillment of the UV-upturn criteria of the NUV-$r$ and FUV-NUV colors. In contrast, when utilizing the FUV-$r$ color alone as the definition or measurement, the dependence on metallicity becomes much weaker, but not vanished. Consequently, although the formation of EHB stars is not directly influenced by metallicity, the resulting UV-upturn phenomenon is. A similar dependence on metallicity within binary star models was also highlighted by \cite{HB14}.

\subsection{The role of dust attenuation}
\label{section:discussion-dust}

Although elliptical galaxies are generally found to be quenched and dust-free \citep[e.g.][]{Rettura06,Rettura11,Barber07,Ali18a}, a considerable fraction of ETGs exhibit RSF activities and internal dust extinction (e.g. $27\%$ in \citealt{Yi11}, $\sim30\%$ in \citealt{Kaviraj07}, $\sim10\%$ in \citealt{Donas07}, and $17\%$ in \citealt{Werle20}), depending on their different sample selection criteria and definitions of RSF. With a photometric color selection for relatively red galaxies, such RSF activities cannot be totally removed. As demonstrated in \cite{Phillipps20}, optical red sequence galaxies defined on the (g-r) versus $M_{\rm r}$ color magnitude diagram are significantly contaminated by RSF galaxies, which can be distinguished by their NUV-$r$ colors. Some of these RSF galaxies are indeed ETGs \citep[e.g.][]{Salim10}, while some of them are in fact red spiral galaxies \citep[e.g.][]{Sodre13,Crossett14,Mahajan20}. Even with a relatively strict additional color criterion of NUV-$r$ $> 5.4$, RSF interlopers with strong dust attenuation still exist as discussed in~\autoref{section:dust}, and could contribute $20\%-60\%$ of the UV-upturn galaxies depending on different dust attenuation models.

In previous studies which focused on UV-upturn phenomenon in old stellar populations, galaxies with RSF activities were treated carefully with many different methods, such as morphological sample selection \citep{Ree07}, optical to mid-infrared SED fitting \citep{Atlee09}, emission line features \citep{Loubser11}, additional star formation sensitive colors \citep[{\sl WISE} W2-W3,][]{Phillipps20}, and image stacks to marginalize their effect \citep{DePropris22}. With either method, it is crucial to carefully exclude such RSF interlopers in the discussion of the genuine UV upturn in old stellar populations. In the upper left panels of~\autoref{fig:Fb_Fy} and~\autoref{fig:Fb_metal}, we demonstrate that such dust-induced UV-upturn galaxies exhibit a strong preference for young stellar age and high stellar metallicity. Consequently, without a clean sample selection, additional unexpected age and metallicity dependencies could be introduced by these RSF interlopers.

\subsection{Limitations and outlook}
\label{section:discussion-limits}

Utilizing the {\Yehb} EPS model, our semi-analytic model {\GABE} roughly reproduced the observed bimodal color distributions of elliptical galaxies. However, it yields a much narrower FUV-NUV range and an inaccurately positioned star-forming sequence, as shown in~\autoref{fig:SAM}. Achieving a fully reproduction of the observed color distributions of galaxies remains a difficult challenge in semi-analytic models. Firstly, galaxy colors are generally not used as a model calibrator. Instead, the stellar mass function is the predominant observational result employed for calibration in semi-analytic models \citep[e.g.][]{Guo11}. Furthermore, scaling relations, such as the supermassive black hole to stellar bulge mass relation, the galaxy size-mass relation, and the metallicity-mass relation, are also ensured during the calibration process. In contrast to these carefully treated galaxy properties, galaxy colors are predictions of semi-analytic models, rather than calibrations. Consequently, semi-analytic models may not perform their best in predicting galaxy colors within the feasible parameter space. Secondly, galaxy colors are highly dependent on the SFHs of galaxies, which are challenging to reproduce self-consistently across various types of galaxies and across different redshifts. Unlike integral galaxy properties such as stellar mass, the star formation rate is differential and sensitive to almost all galactic physical processes, encompassing reionization, hot gas cooling, supernova feedback, galaxy mergers, etc., especially tidal and ram-pressure stripping in satellite galaxies and AGN feedback in massive galaxies. Lastly, numerous physical processes that might influence galaxy colors have not been exquisitely considered in semi-analytic models. As discussed in previous sections, incorporating variations of IMFs, dust attenuation, and binary fractions can significantly alter the distribution of model galaxies on the color-color diagram. The narrower distributions of model galaxies in comparison to the observed one actually reflect such physical processes and their variations in the real Universe which are missed in current semi-analytic models.

In recent years, great efforts have been devoted to better reproduce the SFHs of galaxies, not only on a global scale but also on the satellite scale. For instance, in the GAlaxy Evolution and Assembly (GAEA) model, a gas ejection rate that decreases significantly with cosmic time as suggested by hydrodynamical simulations was incorporated to suppress star formation at high redshift \citep{Hirschmann16}; and a more continuous stripping of hot gas in satellite galaxies was implemented, allowing satellite galaxies to quench in a more gradual way \citep{Xie20}. In the L-GALAXIES semi-analytic model, the star formation in isolated low-mass galaxies is significantly delayed by a longer reincorporation time of ejected gas \citep{Henriques13}; the quenching of satellite galaxies is relieved by only applying ram-pressure stripping in massive halos ($\gtrsim10^{14}\Msun$) and halving the gas surface density threshold for star formation \citep{Henriques17}; and the quenching of $M_{\rm *}$ galaxies is enhanced by increasing the AGN feedback efficiency at lower redshifts \citep{Henriques15}. Both the GAEA and L-GALAXIES models have re-calibrated their model parameters to better reproduce observed stellar mass functions and quenched fractions at $0<z<3$, achieving success even in comparison with the latest observational constraints on quenched fractions from the DESI legacy imaging survey at $0.01\leq z \leq0.08$ \citep[Fig.~13 of][]{Meng23}. The physical recipes of {\GABE} basically follow those of \cite{Guo11}, which are classic but somewhat outdated. We plan to update and re-calibrate our semi-analytic model in our future work, together with a detailed He abundance evolution of galaxies, which is necessary for the He-rich EHB formation model.

Nevertheless, EPS model is crucial or even more determinative in the reproduction of galaxy color distribution, which actually draws the allowed boundary of the color distribution (without considering dust attenuation). Semi-analytic model can only infer galaxy colors within the EPS model-allowed region according to each galaxy's SFH. The galaxy color distributions adopting different EPS models could exhibit significant variation even with the same semi-analytic model, as shown in~\autoref{fig:SAM}. Therefore, in the near future, a well-constructed EPS model with various IMFs and metallicities, which has been observationally constrained by more recent observations \citep[e.g.][]{Sana12,Moe17}, is necessary for the better reproduction of galaxy colors and would be a powerful tool to explore the possible binary origin of the UV-upturn phenomenon.

\section{Conclusions}
\label{section:conclusion}

UV-upturn galaxies are elliptical galaxies (or central bulges in disk galaxies) with abnormally excessive flux in the FUV band. Although the origin of such UV-upturn phenomenon is still under debate, it is generally ascribed to the formation of EHB stars and their hot descendants in old stellar populations (See \autoref{section:intro} for a brief summary of the literature). By using the semi-analytic model of galaxy formation and evolution {\GABE} \citep{Jiang19}, we embed the binary star formation model of EHB stars \citep{Han02,Han03,Han07,Zhang20} within a framework of cosmological evolution to introduce variances of SFHs and explore the formation and evolution of UV-upturn galaxies in a cosmological context. We have gone through all physical mechanisms that are possibly related to the UV-upturn phenomenon, including stellar evolution (EPS models), initial mass function (IMF) and dust attenuation, as well as the age, metallicity and binary fractions of stellar populations in a galaxy, attempting to figure out which processes play a more important role in forming/identifying a UV-upturn galaxy. 

The photometric definition of UV-upturn galaxies of \cite{Yi11} is adopted, and our conclusions can be summarized as follows.
\begin{itemize}
    \item Having examined vaious EPS models including binary and single star versions of {\YunnanII}, {\sc bpass}, as well as {\sc galaxev}, we find that only the {\YunnanII} model which has considered the formation of EHB stars through binary interactions ({\Yehb}) could provide enough FUV flux in old stellar populations and reproduce a relatively large fraction of UV-upturn galaxies. However, all these models are actually not capable of reproducing the entire FUV-NUV range of colors observed in the NSA sample of elliptical galaxies (\autoref{fig:SAM}).
    \item EPS models with various IMFs result in almost identical UV-upturn phenomenon in the semi-analytic model. The variation of the high-mass end slope of IMF has almost no impact on quenched galaxies and only slightly influences the photometric properties of star-forming galaxies.
    \item After considering the dust attenuation, we find two categories of UV-upturn galaxies in the semi-analytic models: old metal-rich quenched elliptical galaxies which are intrinsic UV-upturn galaxies induced by EHB stars in their old stellar populations, and dusty star-forming galaxies which are relatively young galaxies and may also be photometrically identified as UV-upturn galaxies due to dust attenuation. The contribution of the dust attenuation channel depends on the detailed dust attenuation models adopted and ranges from $20\%-60\%$. Dust attenuation curves with strong UV bumps and steep slopes are helpful to the formation of dust induced UV-upturn galaxies.
    \item For the binary star model of \cite{Han07}, we find that the fraction of UV-upturn galaxies induced by EHB stars arising from binary interactions exhibits a strong dependence on both the age and metallicity of galaxies. Old and metal-rich galaxies have greater chances to be UV-upturned.
\end{itemize}

\section*{Acknowledgements}
This work is supported by the National Key R\&D Program of China (grant NO. 2022YFA1602902), the National Natural Science Foundation of China (grant Nos. 12433003, 11821303, 11973030), and China Manned Space Program through its Space Application System.

\section*{Data availability}
The Millennium Simulation data are available at Millennium Database\footnote{\url{https://wwwmpa.mpa-garching.mpg.de/millennium/}}. The simulated galaxies catalogues produced from {\GABE}, and data that support the figures are available from the corresponding author upon reasonable request.



\bibliographystyle{mnras}
\bibliography{UVX}



\appendix

\section{Redshift evolution of UV-optical colors}

\subsection{Dependence on dust attenuation model}
\label{appendix:redshift_dust}

\begin{figure*}
  \centering
  \includegraphics[width=0.85\textwidth]{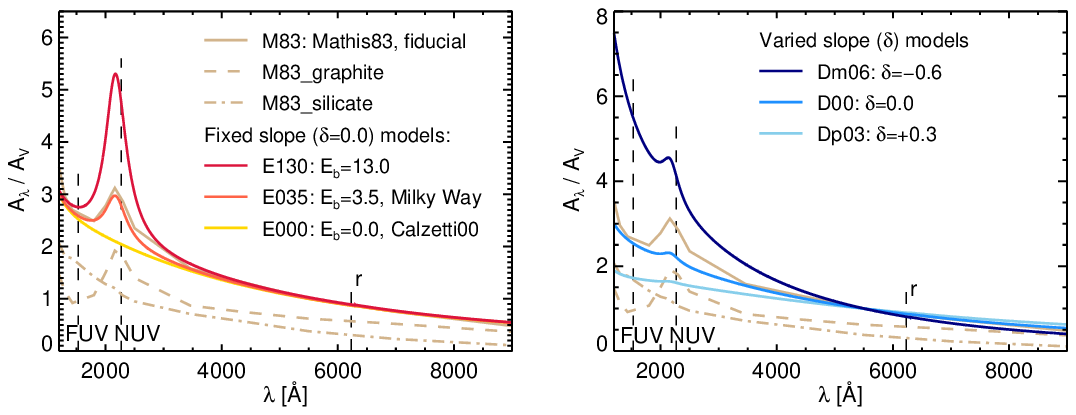}
  \caption{Dust attenuation curves as parameterized in~\autoref{tab:dust_models}. Tan solid line is our fiducial model \protect\citep{Mathis83}; tan dashed and dot-dashed line are the contributions of graphite and silicate respectively. Red and blue solid lines are dust curves with different UV bump amplitude $\Eb$ and slope $\delta$ respectively. Vertical dashed lines mark the effective wavelengths of the FUV (1528 Å), NUV (2271 Å, \protect\citealt{Morrissey05}) and r band (6230 Å, \protect\citealt{SDSS}).}
  \label{fig:dustcurve}
\end{figure*}

\begin{figure*}
  \centering
  \includegraphics[width=0.85\textwidth]{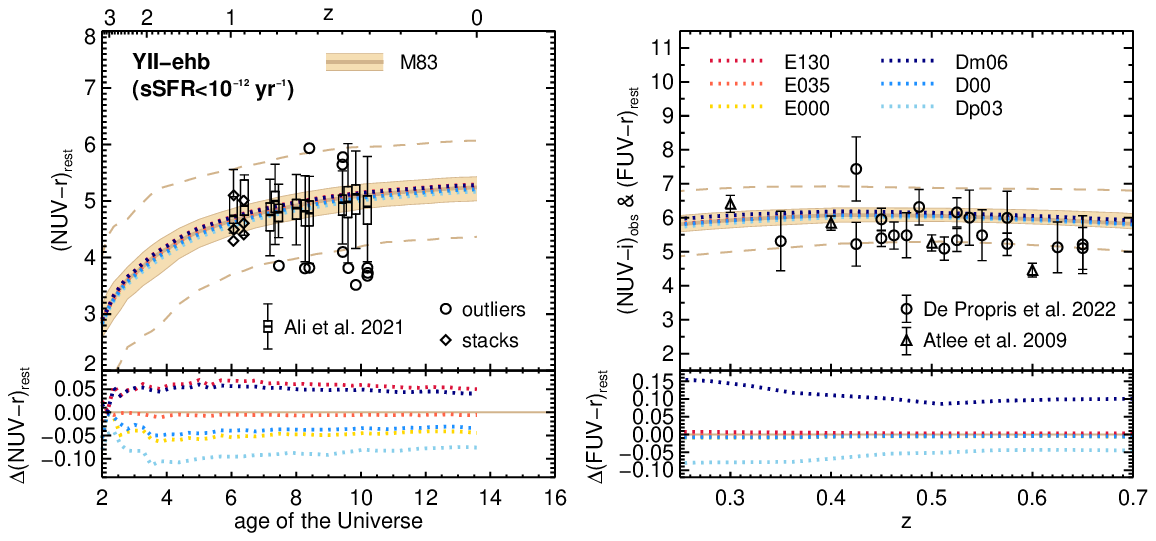}
  \caption{Redshift evolution of the rest-frame NUV-$r$ ({\sl left}) and rest-frame FUV-$r$ colors ({\sl right}) of modeled quenched galaxies (sSFR $<10^{-12}\peryr$) run with the {\Yehb} EPS model and various dust attenuation models as parameterized in~\autoref{tab:dust_models}. Symbols with error bars show the observational measurements, repeated from~\autoref{fig:redshift}. Lower panels show the differences between different dust models and the fiducial M83 model. The tan horizontal line indicates zero.}
  \label{fig:redshift_dust}
\end{figure*}

\begin{figure*}
    \centering
    \includegraphics[width=0.85\textwidth]{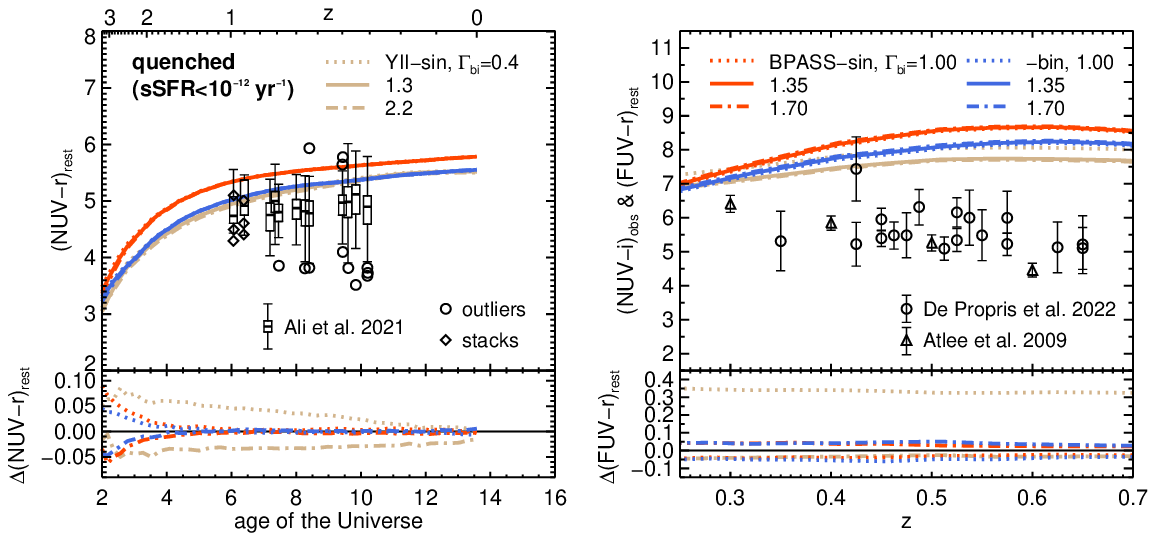}
    \caption{Redshift evolution of the rest-frame NUV-$r$ ({\sl left}) and rest-frame FUV-$r$ colors ({\sl right}) of quenched model galaxies (sSFR $<10^{-12}\peryr$), run with {\Ysin}, {\sl BPASS-sin}, and {\sl BPASS-bin} EPS model with various IMFs. Symbols with error bars show the observational measurements, repeated from~\autoref{fig:redshift}. Lower panels show the differences between different IMFs and the one with $\Gamma_{\rm bi}=$ 1.3 (for {\Ysin}) or 1.35 (for {\sl BPASS-sin} and {\sl BPASS-bin}).}
	\label{fig:redshift_IMF}
\end{figure*}

Dust attenuation curves as parameterized in~\autoref{tab:dust_models} are shown in~\autoref{fig:dustcurve}. As described in~\autoref{section:dust}, in addition to our fiducial curve of \cite{Mathis83}, we consider two sets of models of dust attenuation curves with varying parameters $\delta$ and $\Eb$. The first set (E130, E035, E000) is aimed to isolate the effects of UV bump strength, with a fixed slope of $\delta=0$ but three different values of $\Eb$, as shown in the left panel. The second set of models (Dm06, D00, Dp03) are assumed to follow the observed relation of $\Eb = -1.9\delta + 0.85$ from \cite{Kriek13} to explore the effects of varying slopes with a relatively weak bump, as shown in the right panel. Vertical dashed lines in~\autoref{fig:dustcurve} mark the effective wavelengths of the FUV, NUV, and $r$ band respectively. It is evident that the UV bump strength mainly influences the extinction in the NUV band, applying little effect in the FUV band. With a stronger UV bump, the NUV-$r$ color becomes redder while the FUV-NUV color becomes bluer, which agrees with the behavior of {\UVXdust} galaxies in the E130 model of~\autoref{fig:dust_bump}. In contrast, with a steeper slope, both the FUV-$r$ and NUV-$r$ colors become redder, and the extinction in the FUV band is even stronger than in the NUV band.

\autoref{fig:redshift_dust} shows the redshift evolution of the rest-frame NUV-$r$ (left) and rest-frame FUV-$r$ colors (right) of modeled quenched galaxies (sSFR $<10^{-12}\peryr$) run with the {\Yehb} EPS model and various dust attenuation models as parameterized in~\autoref{tab:dust_models}. The differences between different dust models and the fiducial M83 model are in the lower panels. We find that different dust models merely have a tiny influence on the UV-to-optical colors of these quenched galaxies, with $\Delta$(NUV-$r$) $\lesssim0.10$ and $\Delta$(FUV-$r$) $\lesssim0.15$, as these quenched galaxies are naturally devoid of dust. The differences are much smaller than the observational uncertainties, as indicated by the data points with error bars, hence negligible.

Although the influence is slight, the overall trend agrees with that of~\autoref{fig:dustcurve}. Dust attenuation model with a stronger UV bump (E130) results $\sim0.05$ redder NUV-$r$ colors, while it has almost no influence on the FUV-$r$ colors. In contrast, the one with a steeper slope (Dm06) results both redder NUV-$r$ and FUV-$r$ colors, and the impact on the FUV-$r$ colors is larger.

\subsection{Dependence on initial mass function}
\label{appendix:redshift_IMF}

Similar to \autoref{fig:redshift_dust}, \autoref{fig:redshift_IMF} shows the redshift evolution of the rest-frame NUV-$r$ (left panel) and FUV-$r$ (right panel) colors for modeled quenched galaxies, run with the {\Ysin}, {\sl BPASS-sin}, and {\sl BPASS-bin} EPS model with various IMFs, as indicated. Note that  {\Yehb} EPS models are not included in this analysis for the reason explained in \autoref{section:IMF}. As can be seen from \autoref{fig:redshift_IMF}, all the models considered predict redder UV-to-optical colors than the observation. This is consistent with the conclusion in the main part of the paper that only the {\Yehb} model which has considered the formation of EHB stars through binary interactions could provide enough FUV flux in old stellar populations. Overall, it is clear that the variations in the colors between models of different IMFs are relatively small, less than $\sim0.1$ magnitude in all cases except the {\Ysin} model with $\Gamma=0.4$, which predicts even redder FUV-$r$ colors than the same models  but with steeper IMFs. These results reinforce the conclusion that variations in the high-mass end slope of IMFs have a negligible effect on the UV-upturn phenomenon.


\bsp	
\label{lastpage}
\end{document}